# Conformal Gauge Relativity

## On the Geometrical Unification of Gravitation and Gauge Fields

Eng. Juan Andrés Musante Apolo

Montevideo, Uruguay

Email: ConformalGaugeRelativity@hotmail.com

Web: www.ConformalGaugeRelativity.cubi.ca

### Abstract

A Lagrangian depending on geometric variables (metric, affine connection, gauge group generators) is given which maintains compatibility with General Relativity. It generates the dynamics for Electromagnetism and other Gauge Fields along with Gravitation, at the time it gives a geometric foundation for the stress-energy tensor of continuous matter. The geometric-invariance principle under this integration is exposed and the resulting field equations are obtained. The theory is developed over the tangent space of a four-dimensional real manifold and the generators become those from the Homogenous Lorentz group.

# Contents

# 1 Introduction

Albert Einstein's *General Relativity* is the current theory accepted for describing *Gravitation*. It elegantly explains the gravitational force as the manifestation of space-time curvature caused either by an arbitrary movement of the observer's reference-frame or by the presence of energy (in all its forms, including mass) within its surroundings. Although these causes are essentially different in nature they produce the same effect: they change the geometry perceived. This was clearly expressed by Einstein in his "*Equivalence Principle*" where he stated that inertial and gravitational forces act identically on mass. Free falling objects are deviated by inertial and gravitational forces because they follow geodesics which depart from strait lines when the geometry is curved. Energy, mass and all physical fields evolve over this curved space-time and are responsible for its curvature at the same time. In this elegant picture geometry plays a fundamental role.

Although this has been successfully contrasted in many ways up to a high degree of accuracy, the theory is not complete in the sense that there are many issues for which it doesn't give an explanation. Let's recall some of them:

- Why space-time is four dimensional or seems to be like that?
- Can *matter* be modeled out of some geometric aspect of the manifold?
- Is the *Cosmological Constant* null?
- Can *Electromagnetism* be integrated with *Gravitation* in a geometric way?
- If the previous is possible can other gauge fields be included along with *Electromagnetism*?
- From where the *Standard Model's* forces and symmetries come from?

Mainly those aspects are conveniently fixed or ignored in order to move on to reality modeling but it would be nice if the theory itself could explain them. Of course this doesn't reduce *General Relativity*'s strength but constitutes the very motivation for latter research performed by Einstein and many other mathematicians and physicists like David Hilbert, Ernst Reichenbächer, Jan Schouten, Arthur Eddington, Hermann Weyl, Élie Cartan, Theodor Kaluza and Oskar Klein to mention the very firsts. Unification of all known forces under a single theory where each one can be seen as a different aspect of the existing geometry is by no means one of the most attractive goals in today's physics.

To integrate all known forces into the same picture more facts have to be addressed apart from those covered by *General Relativity*: the *Standard Model* describing particle physics, the mechanism by which particles gain the known masses, the internal symmetries in play, the existing particle flavors and families, not to mention the inevitable *Quantum Mechanics* necessary to understand and work with all of them. Such things can be observed in the labs and mostly can be correctly explained with the actual quantum models developed for describing particle physics.

No matter how complex results the model for describing particle and field interactions or how incompatible *Quantum Mechanics* and *General Relativity* appear to be: it is highly desirable that the theory beyond could be mainly based on and ruled out by the geometric properties of space-time and the internal-space. Since quantum modeling of reality is the correct way to describe things at very short scales and *General Relativity* should be its classical limit at the standard scale then a geometric foundation for a unified quantum field theory would be responsible for the geometric structure described by Einstein's theory. That's why it is natural to think on quantizing the gravitational field in order to take its geometrical content down to the quantum world just to have a clue on what a geometrical foundation would look like at that level. But *Gravity* has resisted to every standard quantizing attempt. Actually any *Quantum Mechanics* field theory is being defined over a flat space-time (null connection and flat metric) and introduces gauge fields through the connection and curvature of some associated internal-space. *Gravity* on the other hand is all about space-time's curvature (standard connection and curved metric). Even if both theories were taken to operate using the same connection their Lagrangians would not combine well in their present form without affecting each other. These facts make both approaches quite incompatible from the geometric point of view. What is needed is a common modeling that better integrates the geometry into its foundations.

The present theory achieves such unification working at the classical level and it is assumed that any symmetry or geometric property considered will be reflected on the quantum domain. Starting with expressions as general as possible it will be shown how to build a Lagrangian compatible with *General Relativity* that includes *Electromagnetism* and other gauge fields. The resulting theory requires the manifold to be four-dimensional.

This is a brief description of the topics seen on each chapter:

On chapter 2 the main geometric variables and some of their properties are introduced (that is the metric and the connection; gauge group generators will be introduced latter on chapter 7). A postulate is proposed on how to build the theory's Lagrangian based on them. An alternative to the known Lagrange's equation is also given which results helpful for obtaining the field's equations. Finally a derivative operator is introduced for achieving a more compact notation when expressing the resulting equations.

On chapter 3 it is shown that Einstein's *General Relativity* can be obtained following the previous postulate when using the Einstein-Hilbert Lagrangian. The concept of compatible theory is introduced. Observations are made with respect to those limitations found when trying to introduce additional fields like *Electromagnetism*.

On chapter 4 an alternative Lagrangian is given for obtaining a theory compatible with *General Relativity*. The solution-conditions for achieving such compatibility are given. It is shown how a stress-energy tensor can be obtained out of geometric components. Two solutions are given as examples: one for the vacuum and another representing a relativistic perfect fluid. Some insight is obtained about the nullity of the *Cosmological Constant*.

On chapter 5 the previous Lagrangian is extended for including *Electromagnetism*. The corresponding field equations are obtained as well as the known compatible set. It is explained why currents can't be introduced with the usual action terms.

On chapter 6 the invariance that allows introducing *Electromagnetism* is analyzed in more detail. It is seen how it affects tensors and derivative operators. Rules are given to restrict tensor algebra according to this invariance. Some insight is obtained about the possibility of introducing mass terms based on vector potentials when modeling bosonic contributions on a particle-field Lagrangian.

On chapter 7 it is explained how other gauge fields can be introduced. The corresponding group generators become fields representing the Homogenous Lorentz group and will be taken as fundamental variables like the metric and the connection. The Lagrangian is modeled to grant their known properties and some additional one. An analysis is carried out on how the torsion can be represented by independent internal components and for which internal symmetry are they responsible. Based on that a suggested interpretation is given for matching this scheme with the Standard Model's structure which can be used as a guidance in future research related to the introduction of particle-fields within the current theory. Also a more fundamental set of elementary particles is proposed which may find its place in such extended theory.

On chapter 8 the Lagrangian is given for introducing gauge fields in a way that compatibility is granted on the resulting solution set. Field equations are displayed and a partial sight is given on the equivalent compatible set.

On chapter 9 a resume is done on the conclusions obtained and some suggestions are given for future research based on this paper.

# 2 Geometrical Objects and Postulate

## 2.1 Metric and Generalized Connection

This field theory will be started on a generic $n$-dimensional real manifold $M^n$ containing a *Riemann Metric* $g^{ij}$ which will be used as usual for measuring lengths, angles and any other relevant geometric quantity (distance, surface, volume, etc.):

(2.1.1) $\exists \; g^{ij} \equiv g^{ij}(X) \quad / \quad g^{ij} = g^{ji}$ *Riemann Metric*

(2.1.2) $\exists \; g_{ij}(X) \quad / \quad g^{ik}.g_{kj} = \delta^i_j$

The manifold will also contain a *General Affine Connection* $\tilde{\Gamma}^i{}_{jk}$ which allows performing parallel transport operations between contiguous tangent spaces $T_p(X)$, and the corresponding tensor covariant differentiation. This connection will be in general non-symmetric in its lower index pair.
It will be useful to express it as the sum of a *Symmetric Connection* $\bar{\Gamma}^i{}_{jk}$ and a *Torsion* tensor $\hat{\Gamma}^i{}_{jk}$:

(2.1.3) $\exists \; \tilde{\Gamma}^i{}_{jk} \equiv \bar{\Gamma}^i{}_{jk} + \hat{\Gamma}^i{}_{jk} \quad / \quad \bar{\Gamma}^i{}_{jk} = \bar{\Gamma}^i{}_{kj}$ *Symmetric-Connection*

(2.1.5) $\hat{\Gamma}^i{}_{jk} = -\hat{\Gamma}^i{}_{kj}$ *Torsion*

(2.1.6) $\forall \; V^i(X) \; , \; W_i(X) \rightarrow \begin{cases} \tilde{\nabla}_j V^i \equiv \partial_j V^i + \tilde{\Gamma}^i{}_{kj}.V^k \\ \tilde{\nabla}_j W_i \equiv \partial_j W_i - \tilde{\Gamma}^k{}_{ij}.W_k \end{cases}$ *Covariant Derivative*

Notice that this *Torsion* is half of the standard torsion tensor:

(2.1.6) $\hat{\Gamma}^i{}_{jk} \equiv \frac{1}{2}.(\tilde{\Gamma}^i{}_{jk} - \tilde{\Gamma}^i{}_{kj}) = \frac{1}{2}.T^i{}_{jk}$

When the so called *Compatibility Condition* is imposed to the metric the *Standard Riemann Connection* can be obtained from it and its first derivatives:

(2.1.7) $\nabla_k g^{ij} = \partial_k g^{ij} + \Gamma^i{}_{rk}.g^{rj} + \Gamma^j{}_{rk}.g^{ir} \equiv 0$ *Compatibility Condition*

(2.1.8) $\rightarrow \quad \Gamma^k{}_{ij} = \frac{1}{2}.g^{kr}.\{\partial_r g_{ij}\} \equiv \frac{1}{2}.g^{kr}.(\partial_i g_{rj} + \partial_j g_{ir} - \partial_r g_{ij})$ *Standard Connection*

The general connection can be expressed as the sum of the standard one plus a *Delta Tensor* $\Delta^i{}_{jk}$ which in turn can be decomposed into a *Symmetric-Delta* tensor and the *Torsion*:

(2.1.9) $\tilde{\Gamma}^i{}_{jk} = \Gamma^i{}_{jk} + \Delta^i{}_{jk} = \Gamma^i{}_{jk} + \bar{\Delta}^i{}_{jk} + \hat{\Delta}^i{}_{jk} \quad / \quad \begin{cases} \bar{\Delta}^i{}_{jk} = \bar{\Delta}^i{}_{kj} & \textit{Symmetric-Delta tensor} \\ \hat{\Delta}^i{}_{jk} = \hat{\Gamma}^i{}_{jk} & \textit{Torsion} \end{cases}$

## 2.2 Connection's derived tensors

The following vector fields can be defined from the general connection by self contractions:

(2.2.1) $\bar{\Gamma}_i \equiv \bar{\Gamma}^k{}_{ik}$ *Symmetric-Connection Trace*

(2.2.2) $\hat{\Gamma}_i \equiv \hat{\Gamma}^k{}_{ik}$ *Torsion Trace*

(2.2.3) $\overset{\rightharpoonup}{\tilde{\Gamma}}_i \equiv \tilde{\Gamma}^k{}_{ik} = \bar{\Gamma}_i + \hat{\Gamma}_i$ *Right-Connection Trace*

(2.2.4) $\overset{\leftharpoonup}{\tilde{\Gamma}}_j \equiv \tilde{\Gamma}^k{}_{kj} = \bar{\Gamma}_j - \hat{\Gamma}_j$ *Left-Connection Trace*

Also the trace of the standard connection can be identified with the following metric's gradient:

(2.2.5) $\Gamma_i \equiv \Gamma^k{}_{ik} = \partial_i(\ln(\sqrt{-|g_{\bullet\bullet}|}))$

With the general connection $\tilde{\Gamma}^k{}_{ij}$ and its first-order derivatives a *Curvature* tensor can be defined along with two related ones obtained from it by self-contraction:

(2.2.6) $\tilde{R}^i{}_{jkl} \equiv \partial_k\tilde{\Gamma}^i{}_{jl} - \partial_l\tilde{\Gamma}^i{}_{jk} + \tilde{\Gamma}^i{}_{rk}.\tilde{\Gamma}^r{}_{jl} - \tilde{\Gamma}^i{}_{rl}.\tilde{\Gamma}^r{}_{jk}$ *Curvature*

(2.2.7) $\overset{\rightharpoonup}{\tilde{R}}_{jl} \equiv \tilde{R}^k{}_{jkl} = \partial_k\tilde{\Gamma}^k{}_{jl} - \partial_l\overset{\rightharpoonup}{\tilde{\Gamma}}_j + \overset{\rightharpoonup}{\tilde{\Gamma}}_r.\tilde{\Gamma}^r{}_{jl} - \tilde{\Gamma}^k{}_{rl}.\tilde{\Gamma}^r{}_{jk}$ *Ricci Tensor*

(2.2.8) $\overset{\leftharpoonup}{\tilde{R}}_{kl} \equiv \tilde{R}^j{}_{jkl} = \partial_k\overset{\leftharpoonup}{\tilde{\Gamma}}_l - \partial_l\overset{\leftharpoonup}{\tilde{\Gamma}}_k$ *Segmental Tensor*

By construction the following identities hold:

(2.2.9) $\tilde{R}^i{}_{jkl} = -\tilde{R}^i{}_{jlk}$

(2.2.10) $\tilde{R}^i{}_{jkl} + \tilde{R}^i{}_{klj} + \tilde{R}^i{}_{ljk} = 2.(\tilde{\nabla}_k\hat{\Gamma}^i{}_{jl} + \tilde{\nabla}_j\hat{\Gamma}^i{}_{lk} + \tilde{\nabla}_l\hat{\Gamma}^i{}_{kj}) + 4.(\hat{\Gamma}^i{}_{kr}.\hat{\Gamma}^r{}_{jl} + \hat{\Gamma}^i{}_{jr}.\hat{\Gamma}^r{}_{lk} + \hat{\Gamma}^i{}_{lr}.\hat{\Gamma}^r{}_{kj})$

(2.2.11) $\tilde{\nabla}_m\tilde{R}^i{}_{jkl} + \tilde{\nabla}_k\tilde{R}^i{}_{jlm} + \tilde{\nabla}_l\tilde{R}^i{}_{jmk} = -2.(\tilde{R}^i{}_{jkr}.\hat{\Gamma}^r{}_{lm} + \tilde{R}^i{}_{jlr}.\hat{\Gamma}^r{}_{mk} + \tilde{R}^i{}_{jmr}.\hat{\Gamma}^r{}_{kl})$

(2.2.12) $\overset{\rightharpoonup}{\tilde{R}}_{jl} = \partial_k\tilde{\Gamma}^k{}_{jl} - \tilde{\Gamma}^r{}_{jk}.\tilde{\Gamma}^k{}_{rl} - \tilde{\nabla}_l\overset{\rightharpoonup}{\tilde{\Gamma}}_j$

(2.2.13) $\overset{\leftharpoonup}{\tilde{R}}_{ij} = -\overset{\leftharpoonup}{\tilde{R}}_{ji}$

Curvatures corresponding to different connections defined over the same manifold (that is connections differing on a generic tensor $D^i{}_{jk}$) become related by the following expressions:

(2.2.14) $\widehat{\Gamma}^i{}_{jk} \equiv \tilde{\Gamma}^i{}_{jk} + D^i{}_{jk}$ $\downarrow$

(2.2.15) $\widehat{R}^i{}_{jkl} = \partial_k\widehat{\Gamma}^i{}_{jl} - \partial_l\widehat{\Gamma}^i{}_{jk} + \widehat{\Gamma}^i{}_{rk}.\widehat{\Gamma}^r{}_{jl} - \widehat{\Gamma}^i{}_{rl}.\widehat{\Gamma}^r{}_{jk} =$

$$= \tilde{R}^i{}_{jkl} + \tilde{\nabla}_k D^i{}_{jl} - \tilde{\nabla}_l D^i{}_{jk} - 2.\hat{\Gamma}^r{}_{kl}.D^i{}_{jr} + D^i{}_{rk}.D^r{}_{jl} - D^i{}_{rl}.D^r{}_{jk} =$$

$$= \tilde{R}^i{}_{jkl} + \widehat{\nabla}_k D^i{}_{jl} - \widehat{\nabla}_l D^i{}_{jk} - (2.\hat{\Gamma}^r{}_{kl} + D^r{}_{kl} - D^r{}_{lk}).D^i{}_{jr} - D^i{}_{rk}.D^r{}_{jl} + D^i{}_{rl}.D^r{}_{jk}$$

Let's introduce a compact notation for the rotational of any vector, shown with the previous traces:

(2.2.16) $\partial\hat{\Gamma}_{ij} \equiv \partial_i\hat{\Gamma}_j - \partial_j\hat{\Gamma}_i$

(2.2.17) $\partial\overset{\leftharpoonup}{\Gamma}_{ij} \equiv \partial_i\overset{\leftharpoonup}{\Gamma}_j - \partial_j\overset{\leftharpoonup}{\Gamma}_i$

(2.2.18) $\partial\overset{\rightharpoonup}{\Gamma}_{ij} \equiv \partial_i\overset{\rightharpoonup}{\Gamma}_j - \partial_j\overset{\rightharpoonup}{\Gamma}_i$

Also some symbols will be defined for the components of the right-contracted curvature and the associated contraction against the metric:

(2.2.19) $\overline{\overset{\rightharpoonup}{\tilde{R}}}_{ij} \equiv \tfrac{1}{2}.(\overset{\rightharpoonup}{\tilde{R}}_{ij} + \overset{\rightharpoonup}{\tilde{R}}_{ji})$ *Symmetric Ricci Tensor*

(2.2.20) $\hat{\overset{\rightharpoonup}{\tilde{R}}}_{ij} \equiv \tfrac{1}{2}.(\overset{\rightharpoonup}{\tilde{R}}_{ij} - \overset{\rightharpoonup}{\tilde{R}}_{ji}) = \tilde{\nabla}_k\hat{\Gamma}^k{}_{ij} + \tfrac{1}{2}.\partial\overset{\rightharpoonup}{\tilde{\Gamma}}_{ij}$ *Skew-Symmetric Ricci Tensor*

(2.2.21) $\overline{\overset{\rightharpoonup}{\tilde{R}}} \equiv g^{kr}.\overline{\overset{\rightharpoonup}{\tilde{R}}}_{kr}$ *Generalized Curvature Scalar*

From (2.2.20) and considering (2.2.3), (2.2.4) and (2.2.8) the following identity can be obtained:

(2.2.22) $\hat{\overset{\rightharpoonup}{\tilde{R}}}_{ij} - \tfrac{1}{2}.\overset{\leftharpoonup}{\tilde{R}}_{ij} = \tilde{\nabla}_k\hat{\Gamma}^k{}_{ij} + \partial\hat{\Gamma}_{ij}$

## 2.3 The Geometrical Postulate

*Geometry* will be imposed into the theory by considering the fundamental geometric objects (the metric, the generalized connection and later the gauge group generators) as dynamic variables which vary independently from one another. Their usage and variational independence comes from the fact that they represent conceptually independent attributes characterizing the containing manifold. Other dynamic variables will be allowed as far as they introduce geometrical constraints on the previous ones (i.e. index symmetries, differential relations, etc.). The Lagrangian will only contain first derivatives of those fundamental variables combined up to conforming second order differential terms:

(2.3.1) *Geometrical Theory* $\rightarrow$ $$I \equiv k.\int_D L(g^{\bullet\bullet},\partial_\bullet g^{\bullet\bullet},\tilde{\Gamma}^{\bullet}{}_{\bullet\bullet},\partial_\bullet\tilde{\Gamma}^{\bullet}{}_{\bullet\bullet},\underset{(x)}{E}{}^{\bullet}{}_{\bullet},\partial_\bullet \underset{(x)}{E}{}^{\bullet}{}_{\bullet},\underset{(y)}{\tilde{C}}).d\Omega$$

(2.3.2) $g^{\bullet\bullet}(X)$ , $\tilde{\Gamma}^{\bullet}{}_{\bullet\bullet}(X)$ , $\underset{(x)}{E}{}^{\bullet}{}_{\bullet}(X)$ $\rightarrow$ *Fundamental Variables*

(2.3.3) / $\underset{(y)}{\tilde{C}}(X)$ $\rightarrow$ *Constraining Variables*

(2.3.4) $$L \equiv \alpha(g^{\bullet\bullet},\partial_\bullet g^{\bullet\bullet},\tilde{\Gamma}^{\bullet}{}_{\bullet\bullet},\partial_\bullet\tilde{\Gamma}^{\bullet}{}_{\bullet\bullet},\underset{(x)}{E}{}^{\bullet}{}_{\bullet},\partial_\bullet \underset{(x)}{E}{}^{\bullet}{}_{\bullet}) + \underset{(y)}{\tilde{C}}.\underset{(y)}{\beta}(g^{\bullet\bullet},\partial_\bullet g^{\bullet\bullet},\tilde{\Gamma}^{\bullet}{}_{\bullet\bullet},\partial_\bullet\tilde{\Gamma}^{\bullet}{}_{\bullet\bullet},\underset{(x)}{E}{}^{\bullet}{}_{\bullet},\partial_\bullet \underset{(x)}{E}{}^{\bullet}{}_{\bullet})$$

(2.3.5) $\rightarrow$ $\delta \underset{(y)}{\tilde{C}}$ ) $\underset{(y)}{\beta}(g^{\bullet\bullet},\partial_\bullet g^{\bullet\bullet},\tilde{\Gamma}^{\bullet}{}_{\bullet\bullet},\partial_\bullet\tilde{\Gamma}^{\bullet}{}_{\bullet\bullet},\underset{(x)}{E}{}^{\bullet}{}_{\bullet},\partial_\bullet \underset{(x)}{E}{}^{\bullet}{}_{\bullet}) \equiv 0$ *Constraining Equations*

## 2.4 Variational Principle and Noëther Currents

The *Variational Principle* states that the action integral must remain stationary with field variations having fixed values on any hypervolume-domain's boundary. This leads to the usual Lagrange's equations:

(2.4.1) $I \equiv k.\int_D L(\underset{(x)}{q},\partial_\bullet \underset{(x)}{q}).d\Omega$ / $\underset{(x)}{q}$ , $\partial_\bullet \underset{(x)}{q}$ *Generic fields and first order derivatives*

(2.4.2) $$\forall \delta \underset{(x)}{q} \, , \; D \;/\; \delta \underset{(x)}{q}(\partial D) \equiv 0 \;\rightarrow\; \left\{ \;\delta I \equiv 0 \;\leftrightarrow\; \partial_k \left( \frac{\partial L}{\partial \partial_k \underset{(x)}{q}} \right) - \frac{\partial L}{\partial \underset{(x)}{q}} = 0 \right.$$

If the action admits symmetries such that remains stationary under certain parameter-variations then by Noëther's theorem a conserved current can be defined based on them:

(2.4.3) $$\forall \delta \underset{(x)}{q} \;/\; \delta I \equiv 0 \;\rightarrow\; \left\{ \;\partial_k J^k = 0 \;/\; J^i \equiv \frac{\partial L}{\partial \partial_i \underset{(x)}{q}}.\delta \underset{(x)}{q} \right.$$

This last equation turns into a charge-conservation law if the corresponding charge is defined as usual:

(2.4.4) $$\forall \partial D \;\rightarrow\; \Delta Q \equiv \oint_{\partial D} J^k.d\Omega_k = 0$$

Because of the *Geometrical Postulate* the constraining variables won't generate any current at all so no physical conserved attribute will be associated to them.

The Lagrangian function $L(\underset{(x)}{q},\partial_\bullet \underset{(x)}{q})$ is a scalar-density so the resulting field equations obtained by using the standard equation (2.4.2) will inevitably involve tensor-densities.

The following method avoids working with them: let's assume the Lagrangian density can be represented as the product of a *Reference Density* $\sigma$ (which can never be null) and a *Lagrange Scalar* function $\Lambda$, both depending on the dynamic variables:

(2.4.5) $$L(\underset{(x)}{q},\partial_\bullet \underset{(x)}{q}) \equiv \sigma(\underset{(x)}{q},\partial_\bullet \underset{(x)}{q}).\Lambda(\underset{(x)}{q},\partial_\bullet \underset{(x)}{q}) \quad / \quad \begin{cases} \sigma(\underset{(x)}{q},\partial_\bullet \underset{(x)}{q}) & \textit{Reference Density} \\ \Lambda(\underset{(x)}{q},\partial_\bullet \underset{(x)}{q}) & \textit{Lagrange Scalar} \end{cases}$$

Based on them the Lagrange's equations can be modified for directly returning tensor expressions:

(2.4.6) *Equivalent Lagrange´s Equations*

$$\partial_k\left(\frac{\partial\Lambda}{\partial\partial_k \underset{(x)}{q}}\right)-\frac{\partial\Lambda}{\partial \underset{(x)}{q}}+\left(\partial_k(\frac{\partial\psi}{\partial\partial_k \underset{(x)}{q}})-\frac{\partial\psi}{\partial \underset{(x)}{q}}+\frac{\partial\psi}{\partial\partial_k \underset{(x)}{q}}.\partial_k\psi\right).\Lambda+\frac{\partial\psi}{\partial\partial_k \underset{(x)}{q}}.\partial_k\Lambda+\frac{\partial\Lambda}{\partial\partial_k \underset{(x)}{q}}.\partial_k\psi=0$$

(2.4.7) $/ \qquad \psi \equiv \ln(\sigma) \quad , \quad \sigma \neq 0$ *Density´s Function*

Also Noëther's currents can be rewritten so that the conservation divergence-law becomes tensorial:

(2.4.8) $$\tilde{\nabla}_k J'^k - (\tilde{\bar{\bar{\Gamma}}}_k - \partial_k\psi).J'^k = 0$$

(2.4.9) $$/ \qquad J'^i \equiv \left(\frac{\partial\Lambda}{\partial\partial_i \underset{(x)}{q}} + \Lambda.\frac{\partial\psi}{\partial\partial_i \underset{(x)}{q}}\right).\delta \underset{(x)}{q}$$ *Equivalent Nöether's Current*

### 2.5 Variational Derivative

Given a connection let's define its *Variational Derivative* by subtracting to the corresponding covariant-derivative operator a term containing the difference between the connection right-trace and the density-function's gradient. For a general connection this will be of the form:

(2.5.1) $$\tilde{\Pi}_k V^i \equiv \tilde{\nabla}_k V^i - (\tilde{\bar{\bar{\Gamma}}}_k - \partial_k\psi).V^i$$ *Variational Derivative*

(2.5.2) $$\tilde{\Pi}_k W^{ij} \equiv \tilde{\nabla}_k W^{ij} - (\tilde{\bar{\bar{\Gamma}}}_k - \partial_k\psi).W^{ij}$$

….

This derivative operator appears naturally in those field equations obtained when the varying parameter of equation (2.4.6) is the generalized connection $\tilde{\Gamma}^i{}_{jk}$, and also in the current's divergence law (2.4.8):

(2.5.3) $$\tilde{\nabla}_k J'^k - (\tilde{\bar{\bar{\Gamma}}}_k - \partial_k\psi).J'^k = 0 \quad \rightarrow \quad \tilde{\Pi}_k J'^k = 0$$

Notice that because of the term subtracted this operator does not follow the Leibnitz rule. When the connection becomes standard this operator reduces to the usual covariant-derivative and the Leibnitz rule is recovered:

(2.5.4) $$\tilde{\Gamma}^i{}_{jk} \equiv \Gamma^i{}_{jk} \quad \rightarrow \quad \tilde{\Pi}_k V^i = \nabla_k V^i$$

# 3 Einstein's Gravitation

## 3.1 Einstein-Hilbert Action

As a starting point let's see how to obtain Einstein's *Gravity* equations out of an action that follows the *Geometrical Postulate*. Since the resulting connection is expected to be symmetric in its lower index pair such condition must be imposed through a suitable Lagrangian:

(3.1.1) $$I_G \equiv -\tfrac{c^3}{16.\pi.k_g}.\int_D \sqrt{-|g_{\bullet\bullet}|}.\left( g^{kr}.\overline{\overline{\tilde{R}}}_{kr} - 2.\lambda_c + \tilde{C}_k^{\ rs}.\hat{\Gamma}^k{}_{rs} \right).d\Omega$$ *Einstein-Hilbert Action*

(3.1.2) / $$\tilde{C}_k^{\ ij}(X) = -\tilde{C}_k^{\ ji}(X)$$ *Lagrange's Multiplier*

The action is written in terms of the metric and the generalized connection as the only fundamental variables. The first two terms in the Lagrangian are those from the Einstein-Hilbert action with a cosmological term. The third one is a Lagrange's multiplier which forces the torsion to be null.
On this action a reference density and a Lagrange's scalar can be easily identified:

(3.1.3) $$\sigma(g_{\bullet\bullet}) \equiv \sqrt{-|g_{\bullet\bullet}|}$$

(3.1.4) $$\Lambda(g^{\bullet\bullet}, \tilde{\Gamma}^{\bullet}{}_{\bullet\bullet}, \partial_{\bullet}\tilde{\Gamma}^{\bullet}{}_{\bullet\bullet}, \tilde{C}_{\bullet}^{\ \bullet\bullet}) \equiv g^{kr}.\overline{\overline{\tilde{R}}}_{kr} - 2.\lambda_c + \tilde{C}_k^{\ rs}.\hat{\Gamma}^k{}_{rs}$$

The resulting density's function has the following non-null derivatives:

(3.1.5) $$\psi \equiv \ln(\sqrt{-|g_{\bullet\bullet}|}) \quad \rightarrow \quad \frac{\partial \psi}{\partial g^{ij}} = -\tfrac{1}{2}.g_{ij} \quad , \quad \partial_i \psi = \Gamma_i$$

The equivalent Lagrange's equations for this kind of action become:

(3.1.6) $$\partial_k \left( \frac{\partial \Lambda}{\partial \partial_k \underset{(x)}{q}} \right) - \frac{\partial \Lambda}{\partial \underset{(x)}{q}} + \Gamma_k.\frac{\partial \Lambda}{\partial \partial_k \underset{(x)}{q}} = 0 \quad / \quad \underset{(x)}{q} \neq g^{ij}$$

(3.1.7) $$\frac{\partial \Lambda}{\partial g^{ij}} - \tfrac{1}{2}.g_{ij}.\Lambda = 0 \quad / \quad \underset{(x)}{q} = g^{ij}$$

## 3.2 Field Equations

The field equations obtained for the action (3.1.1) are:

(3.2.1) $$\delta \tilde{C}_i^{\ jk}\ ) \qquad \hat{\Gamma}^i{}_{jk} = 0$$

(3.2.2) $$\delta g^{ij}\ ) \qquad \overline{\overline{\tilde{R}}}_{ij} - \tfrac{1}{2}.(\overline{\overline{\tilde{R}}} - 2.\lambda_c + \tilde{C}_k^{\ rs}.\hat{\Gamma}^k{}_{rs}).g_{ij} = 0$$

(3.2.3) $$\delta \tilde{\Gamma}^i{}_{jk}\ ) \qquad \tilde{\Pi}_i g^{jk} - \delta_i^k.\tilde{\Pi}_r g^{jr} + 2.\hat{\Gamma}^k{}_{ir}.g^{jr} = \tilde{C}_i^{\ jk}$$

Since the torsion is null this equation set can be further simplified into the following:

(3.2.4) $$\overline{\overline{\tilde{R}}}_{ij} - \tfrac{1}{2}.\overline{\overline{\tilde{R}}}.g_{ij} = -\lambda_c.g_{ij}$$

(3.2.5) → $$\overline{\Pi}_k g^{ij} = 0$$

(3.2.6) $$\tilde{C}_k^{\ ij} = 0$$

Equation (3.2.5) turns into the *Compatibility Condition* if the following connection's extension exists:

(3.2.7) $\forall\ \overline{\Delta}^i{}_{jk} \quad / \quad \overline{\Delta}^j{}_{ri}.g^{rk} + \overline{\Delta}^k{}_{ri}.g^{jr} - \overline{\Delta}^r{}_{ir}.g^{jk} = 0$

(3.2.8) $\widetilde{\Gamma}^i{}_{jk} \equiv \Gamma^i{}_{jk} + \overline{\Delta}^i{}_{jk} \qquad \rightarrow \qquad \nabla_i g^{jk} = 0$

Such extension turns out to be null due to index symmetries so the field equations reduce exactly to *General Relativity* for the vacuum with a given cosmological constant:

(3.2.9) $R_{ij} - \frac{1}{2}.R.g_{ij} = -\lambda_c.g_{ij}$

(3.2.10) $\nabla_i g^{jk} = 0$

With this approach it was shown that *General Relativity* can be obtained by following the proposed *Geometrical Postulate* although the result is limited to the vacuum case. The parameter tensor $\widetilde{C}_i^{jk}$ is a geometric-auxiliary object that imposes an index-symmetry constraint on the resulting connection. It does not participate in the final equations since it cancels to cero.

### 3.3 Compatible Theories

A theory *compatible* with *General Relativity* will be one were the solutions for the later become a particular case in the former solution set. One way for obtaining such a theory from the previous one would be to introduce constraining fields in the action (3.1.1) with terms like the following:

(3.3.1) $I_G \equiv -\frac{c^3}{16.\pi.k_g}.\int_D \sqrt{-|g_{\bullet\bullet}|}.\left( g^{kr}.\overline{\overline{\widetilde{R}}}_{kr} - 2.\lambda_c + \widetilde{C}_k^{rs}.\hat{\Gamma}^k{}_{rs} + \phi \right).d\Omega$

(3.3.2) $/ \qquad \phi \equiv \alpha'(g^{\bullet\bullet}, \partial_\bullet g^{\bullet\bullet}, \widetilde{\Gamma}^\bullet{}_{\bullet\bullet}, \partial_\bullet \widetilde{\Gamma}^\bullet{}_{\bullet\bullet}) + \underset{(y)}{\widetilde{C}}(X).\underset{(y)}{\beta'}(g^{\bullet\bullet}, \partial_\bullet g^{\bullet\bullet}, \widetilde{\Gamma}^\bullet{}_{\bullet\bullet}, \partial_\bullet \widetilde{\Gamma}^\bullet{}_{\bullet\bullet})$

Although this takes to a metric-equation like (3.2.9) with an additional stress-energy tensor derived from the $\phi$ function and its variables, the components $\alpha'$ and $\beta'$ should generate a connection-equation like (3.2.3) which admits the standard connection $\Gamma^i{}_{jk}$ as a solution. This demands the following condition to hold:

(3.3.3) $\underset{(\phi)}{E_i^{jk}} \equiv \partial_r\left( \frac{\partial\phi}{\partial\partial_r \widetilde{\Gamma}^i{}_{jk}} \right) - \frac{\partial\phi}{\partial\widetilde{\Gamma}^i{}_{jk}} + \Gamma_r.\frac{\partial\phi}{\partial\partial_r \widetilde{\Gamma}^i{}_{jk}} \qquad \rightarrow \qquad \underset{(\phi)}{E_i^{jk}} = -\underset{(\phi)}{E_i^{kj}}$

Very restricting is the fact that derivatives are not allowed for the constraining variables making it impossible to obtain second order differential equations on them. This excludes the chance to introduce field-potentials as constraining variables by using derivative terms like with *Electromagnetism*. Since the connection reduces to the standard one the only way to introduce vector potentials would be through the metric by following the Kaluza-Klein construct [8]. This last procedure will be depreciated in favor of a better one.

For overcoming the previous limitations another compatible action will be proposed without the null-torsion constraint. This will allow introducing additional fields (potentials) through the general connection and not through the metric. At the same time it will fix the manifold's dimensionality while enabling a conformal symmetry which will be useful for the theory's interpretation.

# 4 Four-Dimensional Conformal Gravitation

## 4.1 Conformal Gravity Action

By removing the null-torsion constraint and the cosmological term, and by squaring the *Generalized Curvature* scalar in action (3.1.1) the resulting one becomes invariant under conformal transformations when applied to the metric tensor:

(4.1.1) $$I_{CG} \equiv -\tfrac{c^3}{16.\pi.k_g}.\int_D \sqrt{-|g_{\bullet\bullet}|}.\left(g^{kr}.\overline{\overline{\tilde{R}}}_{kr}\right)^2.d\Omega$$ *Conformal Gravity Action*

A conformal transformation on the metric can be expressed as:

(4.1.2) $$g'^{ij} \equiv \mathrm{e}^{2.\varphi}.g^{ij} \qquad , \qquad \tilde{\Gamma}'^{i}{}_{jk} \equiv \tilde{\Gamma}^{i}{}_{jk}$$

This action remains invariant only when the manifold is four-dimensional:

(4.1.3) $$I'_{CG} = -\tfrac{c^3}{16.\pi.k_g}.\int_D \sqrt{-|g'_{\bullet\bullet}|}.\left(g'^{kr}.\overline{\overline{\tilde{R}}}'_{kr}\right)^2.d\Omega =$$

$$= -\tfrac{c^3}{16.\pi.k_g}.\int_D \sqrt{-|g_{\bullet\bullet}|}.\mathrm{e}^{\varphi.(4-n)}.\left(g^{kr}.\overline{\overline{\tilde{R}}}_{kr}\right)^2.d\Omega$$

(4.1.4) $$\rightarrow \quad \forall \quad \varphi(X) \quad , \quad I'_{CG} = I_{CG} \quad \leftrightarrow \quad n = 4$$

Under such conditions if a metric is a solution of the resulting field equations then those obtained from it by a conformal transformation will also be solutions. So the metric will be determined up to a conformal factor. A four-dimensional manifold will be assumed from now on in order to allow this symmetry.

## 4.2 Field Equations

The corresponding field equations become:

(4.2.1) $$\delta g^{ij} \;) \qquad \overline{\overline{\tilde{R}}}.\left(\overline{\overline{\tilde{R}}}_{ij} - \tfrac{1}{4}.\overline{\overline{\tilde{R}}}.g_{ij}\right) = 0$$

(4.2.2) $$\delta\tilde{\Gamma}^{i}{}_{jk} \;) \qquad \tilde{\Pi}_i(\overline{\overline{\tilde{R}}}.g^{jk}) - \delta_i^k.\tilde{\Pi}_r(\overline{\overline{\tilde{R}}}.g^{jr}) + 2.\hat{\Gamma}^{k}{}_{ir}.(\overline{\overline{\tilde{R}}}.g^{jr}) = 0$$

fg

## 4.3 Two Solution Sets

Solutions to the previous field equations can be classified in two separate sets which will be called *Compact* and *Standard*:

(4.3.1) $$\overline{\overline{\tilde{R}}} = 0$$ *Compact Set*

---

(4.3.2) $$\overline{\overline{\tilde{R}}} \neq 0$$

(4.3.3) $$\overline{\overline{\tilde{R}}}_{ij} - \tfrac{1}{4}.\overline{\overline{\tilde{R}}}.g_{ij} = 0$$

(4.3.4) $$\tilde{\Pi}_i g^{jk} - \delta_i^k.\tilde{\Pi}_r g^{jr} + 2.\hat{\Gamma}^{k}{}_{ir}.g^{jr} = \delta_i^k.g^{jr}.\partial_r(\ln\overline{\overline{\tilde{R}}}) - g^{jk}.\partial_i(\ln\overline{\overline{\tilde{R}}})$$

*Standard Set*

From equation (4.3.4) the following equality can be deduced:

(4.3.5) $$\overline{\overline{\tilde{R}}}(X) = 4.\lambda_c.e^{\tilde{\phi}(C,X)}$$ *Curvature- Phase Equation*

(4.3.6) / $$\tilde{\phi}(C,X) \equiv -\int_C (g_{sr}.\tilde{\Pi}_k g^{rk} + \tfrac{2}{3}.\hat{\Gamma}_s).dX^s$$ *Global Phase*

The function $\tilde{\phi}(C,X)$ will be called *Global Phase.* It depends on a curve $C$ which ends at the point $X$ where the equation (4.3.5) is being evaluated.

For the compact set the unique field equation just fix one degree of freedom among those existing within the metric and the *Symmetric Ricci* tensor. This allows lots of solutions which can strongly depart from *General Relativity* ones.

### 4.4 The Uniqueness Constraints

The *Curvature-Phase* equation (4.3.5) involves a generic curve $C$ which ends at the point in space where the equation is being evaluated. The generic *Curvature* scalar should be determined by the field equations up to an arbitrary conformal transformation. But the election of the curve has far more degrees of freedom than a simple conformal transformation. This implies both sides of equation (4.3.5) can't compensate each other to leave a unique well defined function. That's why some extra condition should be imposed in order to calculate a unique solution: either a prescription is given for considering only one curve ending at each point or the theory is tailored to depend as less as possible on these curves.
Based on that one choice has to be made between the following two constraints:

(4.4.1) ***Path Dependence Constraint***

For each point $X$ there exists a prescription for defining a single curve $C$ ending at it for which a unique solution can be calculated.

This converts the *Global Phase* into a well defined function which becomes dependent on the family of curves considered. An example of such rule would be the following: define for each point $X_0$ a family of open curves $\{C\}$ starting at it which can reach every point $X$ in a neighborhood around it with a unique parameterization set. This could be for example a family of geodesics starting at $X_0$ (Gaussian normal coordinates) and the neighborhood will extend to the surface of first intersections. For covering the whole Universe many of these patches would have to be considered with the corresponding boundary conditions for preserving the result's continuity when going from one neighborhood to a contiguous one.

(4.4.2) ***Path Independence Constraint***

The *Global Phase* $\tilde{\phi}$ holds the following property:

(4.4.3) $$\exists\ \tilde{\phi}\ \ /\ \ g_{ir}.\tilde{\Pi}_k g^{rk} + \tfrac{2}{3}.\hat{\Gamma}_i \equiv \partial_i \tilde{\phi}$$ *Path Independence Constraint*

With this constraint equation (4.3.5) will depend only on the starting and ending points of the curve considered, assuming that the manifold is a connected one. Then only a prescription for determining the starting point is needed.

## 4.5 GR-Compatible Field Equations

The solutions for the standard set have a special significance: they will take us back on the track of *General Relativity*. For achieving this lets notice that for the following general connection equation (4.3.4) simplifies very close to the standard *Compatibility Condition*:

(4.5.1) $$\forall \quad \tilde{\Gamma}^{i}{}_{jk} \equiv \Gamma'^{i}{}_{jk} + \Delta^{i}{}_{jk} \qquad , \qquad \Gamma'^{i}{}_{jk} \equiv \Gamma'^{i}{}_{kj}$$

(4.5.2) $/$ *Connection Compatibility Condition*

$$\Delta^{j}{}_{ri}.g^{rk} + \Delta^{k}{}_{ri}.g^{jr} - (g_{ir}.\Delta^{r}{}_{sp}.g^{sp} + \tfrac{2}{3}.\hat{\Delta}_i + \bar{\Delta}_i).g^{jk} + \tfrac{2}{3}.\delta^k_i.\hat{\Delta}_r.g^{jr} + 2.\hat{\Delta}^{k}{}_{ir}.g^{jr} \equiv 0$$

(4.5.3) $$(4.5.2)\ ,\ (4.3.4) \quad \rightarrow \quad \nabla'_i g^{jk} - g^{jk}.g_{ir}.\nabla'_p g^{rp} = 0$$

From this last equation the metric *Compatibility Condition* (2.1.7) cannot be deduced even though both equations share the standard connection as a solution:

(4.5.4) $$\forall \quad \Gamma^{i}{}_{jk} \quad / \qquad \nabla_i g^{jk} = 0 \quad \rightarrow \quad \nabla_i g^{jk} - g^{jk}.g_{ir}.\nabla_p g^{rp} = 0$$

This happens due to the conformal invariance of equation (4.3.4), inherited by (4.5.3) but missing in Einstein's equations. The condition (4.5.2) will be called *Connection Compatibility Condition* and can be decomposed into the following two independent ones:

(4.5.5) $$\hat{\Gamma}^{i}{}_{jk} = -\tfrac{1}{3}.(\delta^i_j.\hat{\Gamma}_k - \delta^i_k.\hat{\Gamma}_j)$$

(4.5.6) $$\Delta_i{}^{k}{}_{j} + \Delta^{k}{}_{ij} - \delta^k_i.(\Delta_j{}^{r}{}_{r} + \Delta^{r}{}_{jr}) = 0$$

*Connection Compatibility Conditions* (4.5.5)–(4.5.6)

One more condition is necessary for obtaining compatibility with *General Relativity*:

(4.5.7) $$\overline{\overline{\tilde{R}}} \equiv 4.\lambda_c$$ *Relativistic Conformal Gauge*

This condition will be called *Relativistic Conformal Gauge* and stands for a special choice on the metric's conformal gauge. Since the Lagrangian is invariant under conformal transformations, imposing this condition won't affect the field equations or the physics of their solutions.

With conditions (4.5.5), (4.5.6) and (4.5.7) when identifying $\Gamma'^{i}{}_{jk}$ with $\Gamma^{i}{}_{jk}$, the standard equation set can finally be transformed into:

(4.5.8) $$\hat{\Delta}^{i}{}_{jk} = -\tfrac{1}{3}.(\delta^i_j.\hat{\Delta}_k - \delta^i_k.\hat{\Delta}_j)$$

(4.5.9) $$\Delta_i{}^{k}{}_{j} + \Delta^{k}{}_{ij} - \delta^k_i.(\Delta_j{}^{r}{}_{r} + \Delta^{r}{}_{jr}) = 0$$

(4.5.10) $$\nabla_i g^{jk} = 0$$

(4.5.11) $$R_{ij} - \tfrac{1}{2}.R.g_{ij} = \frac{8.\pi.k_g}{c^4}.T_{ij} - \lambda_c.g_{ij}$$

(4.5.12) $$/ \quad T_{ij} \equiv \frac{c^4}{8.\pi.k_g}.(\overline{\overline{\Delta}}_{ij} - \tfrac{1}{2}.g^{kr}.\overline{\overline{\Delta}}_{kr}.g_{ij})$$

(4.5.13) $$\overline{\overline{\Delta}}_{ij} \equiv -\nabla_k\bar{\Delta}^{k}{}_{ij} + \bar{\Delta}^{k}{}_{ir}.\bar{\Delta}^{r}{}_{jk} - \tfrac{1}{3}.\hat{\Delta}_i.\hat{\Delta}_j + \tfrac{1}{2}.(\nabla_i\bar{\Delta}_j + \nabla_j\bar{\Delta}_i) - \bar{\Delta}^{r}{}_{ij}.\bar{\Delta}_r$$

This is *General Relativity* having a stress-energy tensor and a cosmological term so its solutions are also solutions of the current theory which proves its compatibility.

A stress-energy tensor (4.5.12) is obtained out of components coming from the general connection in (4.5.13) so it can be said such tensor is the manifestation of some geometric aspect of the manifold. It can be used for modeling *continuous-matter* fields based on the *Geometrical Postulate* (…"*some wood out of marble*"…) gaining this way a strong geometrical foundation.

A non-null torsion (4.5.8) is admitted whose trace contributes to the resulting stress-energy tensor in (4.5.13).

## 4.6 The Compatible Family of Curves

According to (4.3.5) and (4.5.7) the *Relativistic Conformal Gauge* (and so the possibility of having compatible well defined solutions) is equivalent to setting the *Global Phase* to cero:

(4.6.1) $$\widetilde{\phi}(C,X) \equiv 0$$

This can be achieved in a variety of ways and the type of family-curve considered becomes relevant. If any curve is to be allowed then the previous condition translates into:

(4.6.2) $$g_{sr}.\widetilde{\Pi}_k g^{rk} + \tfrac{2}{3}.\hat{\Gamma}_s \equiv 0$$

But the family of curves can be constrained resulting in different compatibility conditions.
For example it's possible to consider only closed curves. In such case condition (4.6.1) becomes equivalent to the *Path Independence Constraint* mentioned in (4.4.2):

(4.6.3) $$g_{ir}.\widetilde{\Pi}_k g^{rk} + \tfrac{2}{3}.\hat{\Gamma}_i \equiv \partial_i \widetilde{\phi} \qquad \rightarrow \qquad \widetilde{\phi}(C,X) \equiv \oint_C \partial_k \widetilde{\phi}.dX^k = 0$$

This constraint may also apply to open curves holding:

(4.6.4) $$\widetilde{\phi}(X) \equiv \widetilde{\phi}(X_0)$$

The bigger the family of curves the smaller will be the set of compatible solutions.

## 4.7 Vacuum solutions

Vacuum solutions are obtained for the following *Vacuum Connection*:

(4.7.1) $$\widetilde{\Gamma}^i{}_{jk} \equiv \Gamma^i{}_{jk} + \delta^i_j.a_k$$ *Vacuum Connection*

Where $a_i$ is an arbitrary vector field that will be called *Vacuum Potential*.
Conditions (4.5.8) and (4.5.9) hold for this connection while the stress-energy tensor (4.5.12) becomes null. Equality (4.6.2) is also true so these solutions live in the *Relativistic Conformal Gauge's* domain. The simplified equation set becomes the one for *General Relativity*'s vacuum as in (3.2.9) and (3.2.10):

(4.7.2) $$R_{ij} - \tfrac{1}{2}.R.g_{ij} = -\lambda_c.g_{ij}$$

(4.7.3) $$\nabla_i g^{jk} = 0$$

Connection (4.7.1) is commonly known as being a projective transformation of the standard one.

### 4.8 Perfect Fluid's Stress-Energy Tensor

As an example of a geometric-like matter field let's see how a Killing vector field can generate the standard stress-energy tensor for a perfect fluid:

(4.8.1) $$\forall\ \alpha_i(X) \qquad / \qquad \nabla_i\alpha_j + \nabla_j\alpha_i = 0$$

With this field the following delta-tensor can be constructed for the general connection:

(4.8.2) $$\Delta^i{}_{jk} \equiv i.(g_{jk}.\alpha^i - \delta^i_j.\alpha_k - \delta^i_k.\alpha_j)$$

Such tensor hold conditions (4.5.8) and (4.5.9) as it should be.
The corresponding stress-energy tensor can be derived following equations (4.5.12) and (4.5.13):

(4.8.3) $$T_{ij} = (\varepsilon + p).V_i.V_j - p.g_{ij}$$

(4.8.4) $$V_i = (\alpha^k.\alpha_k)^{-\frac{1}{2}}.\alpha_i \qquad \textit{Velocity}$$

(4.8.5) $$/ \quad \varepsilon = 3.\tfrac{c^4}{8.\pi.k_g}.\alpha^k.\alpha_k \qquad \textit{Mass Density}$$

(4.8.6) $$p = -\tfrac{c^4}{8.\pi.k_g}.\alpha^k.\alpha_k \qquad \textit{Pressure}$$

Where $V^i$ is the fluid's relativistic velocity field corresponding to the direction of vector $\alpha_i$ and $\varepsilon$ and $p$ the mass-density and pressure respectively, both depending on the field's module.
It's interesting to notice that a complex-value connection is needed for modeling matter. This strongly suggests that the current theory may be well extended to the complex domain. The resulting connection (standard plus (4.8.2)) looks similar in shape to the one considered by Weyl in his unification theory [8].

### 4.9 The Cosmological Constant can't be Null

From (4.3.2) and (4.3.5) it can be seen that the cosmological constant becomes associated to the standard solution set which is the one leading to *General Relativity*. Such constant can't be taken to have a cero value because in that case the solution set would become the compact one. Although compatible solutions may exist under this last set they would not be completely determined by the corresponding field equation. So when talking about *General Relativity* as being well defined by the theory the cosmological constant should be assumed to be different than cero:

(4.9.1) $$\textit{General Relativity} \quad \leftrightarrow \quad \lambda_c \neq 0$$

### 4.10 Relation between Standard Differential Geometry and General Relativity

While the metric is being scaled by conformal transformations, distance vectors and the affined connection are not. That's why not all geometric measurements remain invariant under those transformations. For example angles between invariant vectors are conserved but vector modules are not.

(4.10.1) $$g'_{ij} = \mathrm{e}^{-2.\varphi}.g_{ij} \quad , \quad A'^i = A^i \quad , \quad B'^i = B^i$$

(4.10.2) $$|A|^2 = g_{kr}.A^k.A^r \quad \rightarrow \quad |A'|^2 = \mathrm{e}^{-2.\varphi}.|A|^2 \neq |A|^2$$

(4.10.3) $$\cos(A.B) = g_{kr}.A^k.B^r.|A|^{-1}.|B|^{-1} \quad \rightarrow \quad \cos(A'.B') = \cos(A.B)$$

The space-time interval results to be non-invariant and so do the velocity vector:

(4.10.4) $$dS^2 \equiv g_{kr}.dX^k.dX^r \qquad \rightarrow \qquad dS'^2 = \mathrm{e}^{-2.\varphi}.dS^2 \neq dS^2$$

(4.10.5) $$V^i \equiv \frac{dX^i}{dS} \qquad \rightarrow \qquad V'^i = \mathrm{e}^{\varphi}.V^i \neq V^i$$

For such reasons most geometric measures under the actual conformal theory depends on the gauge selected. *General Relativity* physics is recovered after adopting the *Relativistic Conformal Gauge* and so do the standard measurement essentials (up to a constant scaling factor). Here it can be seen that the standard differential geometry becomes a special case on a more elaborated one where the geometric objects defined should be invariant under conformal transformations of the metric. The fact that the geometry simplifies with the *Relativistic Conformal Gauge* is the reason why *General Relativity* becomes a very special (preferred) reference for describing reality. The relation is similar to the one *Special Relativity* maintains with *General Relativity* being the first a special case of the second: when defined on a neighborhood small enough with respect to the existing curvature and an inertial coordinate-system is being used as a reference.

### 4.11 Nöether Currents

The Nöether currents for each geometric variable result:

(4.11.1) $$\delta g^{ij}\ ) \qquad \underset{(g^{\bullet\bullet})}{J}{}^{i} = 0$$

(4.11.2) $$\delta\tilde{\Gamma}^i{}_{jk}\ ) \qquad \underset{(\tilde{\Gamma}^{\bullet}{}_{\bullet\bullet})}{J}{}^{i} = 2.\overline{\vec{\tilde{R}}}.(g^{kr}.\delta\tilde{\Gamma}^i{}_{kr} - g^{ik}.\delta\vec{\tilde{\Gamma}}_k)$$

So far the currents associated to the metric become null which implies that no "metric charge" can be considered. On this respect the fundamental variable behaves as a constraining one. This wouldn't be the case if the action is constrained for example by introducing the following Lagrange's term which depends on the metric and its first order derivatives:

(4.11.3) $$\sqrt{-|g_{\bullet\bullet}|}.\tilde{C}^i{}_{jklm}.(\nabla_i g^{jk} - g^{jk}.g_{ir}.\nabla_s g^{rs}).g^{lm} \qquad / \qquad \Gamma^k{}_{ij} \equiv \tfrac{1}{2}.g^{kr}.\{\partial_r g_{ij}\}$$

Such term turns the *Connection Compatibility Condition* into a field equation.
The payoff for such change would be a theory having non-null metric currents and a stress-energy tensor containing the parameter $\tilde{C}^i{}_{jklm}$ which becomes a real physical field. The solutions for the theory will be stretched in range while the compatibility with *General Relativity* remains untouched. Terms like this one will be considered later on this paper.

## 5. Conformal Electromagnetic Gravitation

So far the *Four-Dimensional Conformal Gravitation* proved to be a theory compatible with *General Relativity*, but something important is missing: it does not contain *Electromagnetism*. By following the *Geometric Postulate* there is no way to introduce such field only with this action and additional constraining terms. As discussed in (3.3) with the Einstein-Hilbert Lagrangian, the solution will be reached by adding new fundamental terms. For knowing what terms to add a good starting point will be to consider the *Vacuum Potential* as a candidate for the *Electromagnetic* one. Doing so prevents the current action term to generate any unreal *Electromagnetic* stress-energy tensor. The good one would be generated by new terms which should preserve the existing compatibility with *General Relativity*. Only this way *Gravitation* and *Electromagnetism* will be able to merge seamlessly.

### 5.1 Conformal Electromagnetic Gravity Action

Let's consider the *Vacuum Connection* plus a compatible symmetric delta-tensor and identify the vacuum potential with the *Electromagnetic* one:

(5.1.1) $$\tilde{\Gamma}^i{}_{jk} \equiv \Gamma^i{}_{jk} + \Theta^i{}_{jk} + c_1.\delta^i_j.A_k$$ *EM-GR Connection*

(5.1.2) $$\Theta^i{}_{jk} \equiv \Theta^i{}_{kj}$$

/

(5.1.3) $$\Theta_i{}^k{}_j + \Theta^k{}_{ij} - \delta^k_i.(\Theta_j{}^r{}_r + \Theta^r{}_{jr}) \equiv 0$$

This will be called the *Electromagnetic Gravity Connection*, where $A_k$ is the usual *Electromagnetic* potential and $c_1$ the corresponding coupling constant. The torsion of such connection depends entirely on the *Electromagnetic* potential:

(5.1.4) $$\hat{\Gamma}^i{}_{jk} = \tfrac{1}{2}.c_1.(\delta^i_j.A_k - \delta^i_k.A_j) = -\tfrac{1}{3}.(\delta^i_j.\hat{\Gamma}_k - \delta^i_k.\hat{\Gamma}_j)$$ *EM-GR Torsion*

Introducing the potential this way is also supported by the fact that the curvature tensor for any connection remains invariant when the connection is modified in the following way:

(5.1.5) $$\forall \quad \lambda(X) \ , \ \tilde{\Gamma}^i{}_{jk} \ \rightarrow \ \{ \quad \tilde{\Gamma}'^i{}_{jk} \equiv \tilde{\Gamma}^i{}_{jk} + \delta^i_j.\partial_k\lambda \quad \rightarrow \quad \tilde{R}'^i{}_{jkl} = \tilde{R}^i{}_{jkl}$$

This change which was called *Lambda Transformation* by Einstein breaks the index-symmetry on any symmetric connection making this property to lose relevance. It motivated him to look for unification between *Gravitation* and *Electromagnetism* using a non-symmetric field theory. Although such theory did not succeed this transformation found to be useful in the current paper.

For the *EM-GR* connection the lambda function can be seen as representing the gauge freedom of *Electromagnetism*. Latter this will be proved to be correct under the appropriate conditions.

For finding the correct *Electromagnetic* terms for the Lagrangian only those connection-derived tensors being invariant under lambda transformations will be considered. This ensures that after constraining the generalized connection to the *EM-GR* one, the correct equations can emerge. There are only two independent tensors that can be formed from the general connection and its first order derivatives which are lambda-invariant:

(5.1.6) $$\tilde{R}^i{}_{jkl} \quad , \quad \partial\hat{\Gamma}_{ij}$$

All other invariant terms can be derived from them by tensor operations.
For preserving the Lagrangian's conformal symmetry the additional terms should have the form:

(5.1.7) $$\sqrt{-|g_{\bullet\bullet}|}.g^{kr}.g^{sl}.X_{krsl}$$

Where $X_{ijkl}$ is a generic expression for quadratic combinations of lambda-invariant tensors obtained from those in (5.1.6). Such combination of first order differential terms will generate the expected second order differential equations for *Electromagnetism*.

Also those quadratic combinations should not break compatibility with *General Relativity* achieved by the current action term. This rule out many combinations leaving the following terms to be combined:

(5.1.8) $$\hat{\tilde{\tilde{R}}}_{ij} \quad , \quad \tilde{\tilde{R}}_{ij} \quad , \quad \tilde{R}^i{}_{jkl} + \tilde{R}^i{}_{ljk} + \tilde{R}^i{}_{klj} \quad , \quad \partial\hat{\Gamma}_{ij}$$

This is consistent with the fact that all of these terms must vanish when constraining to the standard connection in order to preserve compatibility:

(5.1.9) $$\tilde{\Gamma}^i{}_{jk} \equiv \Gamma^i{}_{jk} \quad \rightarrow \quad \hat{\overrightarrow{\tilde{R}}}_{ij} = \overleftarrow{\tilde{R}}_{ij} = \partial\hat{\Gamma}_{ij} = 0 \quad , \quad \tilde{R}^i{}_{jkl} + \tilde{R}^i{}_{ljk} + \tilde{R}^i{}_{klj} = 0$$

By following those rules a first tentative for the action would be:

(5.1.10)

$$I_{General} \equiv k_0.\int_D \sqrt{-|g_{\bullet\bullet}|}.g^{kr}.g^{sl}.\Big( \overline{\overrightarrow{\tilde{R}}}_{kr}.\overline{\overrightarrow{\tilde{R}}}_{sl} + k_1.\hat{\overrightarrow{\tilde{R}}}_{ks}.\hat{\overrightarrow{\tilde{R}}}_{rl} + k_2.\overleftarrow{\tilde{R}}_{ks}.\overleftarrow{\tilde{R}}_{rl} + k_3.\partial\hat{\Gamma}_{ks}.\partial\hat{\Gamma}_{rl} +$$
$$+ k_4.\hat{\overrightarrow{\tilde{R}}}_{ks}.\overleftarrow{\tilde{R}}_{rl} + k_5.\hat{\overrightarrow{\tilde{R}}}_{ks}.\partial\hat{\Gamma}_{rl} + k_6.\overleftarrow{\tilde{R}}_{ks}.\partial\hat{\Gamma}_{rl} +$$
$$+ k_7.(\tilde{R}^m{}_{kso} + \tilde{R}^m{}_{oks} + \tilde{R}^m{}_{sok}).(\tilde{R}^o{}_{rlm} + \tilde{R}^o{}_{mrl} + \tilde{R}^o{}_{lmr}) +$$
$$+ k_8.g^{mn}.g_{op}.(\tilde{R}^o{}_{ksm} + \tilde{R}^o{}_{mks} + \tilde{R}^o{}_{smk}).(\tilde{R}^p{}_{rln} + \tilde{R}^p{}_{nrl} + \tilde{R}^p{}_{lnr}) \Big).d\Omega$$

Within this action some extra conditions called *EM-GR Splitting Conditions* must be met by the constants $k_\bullet$ for preserving compatibility and obtaining a neat separation between *Gravity* and *Electromagnetism*:

(5.1.11) $$\left. \begin{array}{lcl} k_2 = \frac{1}{4}.k_1 & , & k_4 = -k_1 \\ k_6 = -\frac{1}{2}.k_5 & , & k_7 = -\frac{1}{2}.k_1 - \frac{3}{8}.k_5 - 3.k_8 \end{array} \right\} \quad \textit{EM-GR Splitting Conditions}$$

(5.1.12) $$\delta\tilde{\Gamma}^\bullet{}_{\bullet\bullet} \text{ - field equation} \quad \rightarrow \quad \nabla_i g^{jk} = 0 \quad , \quad \nabla_k F^{jk} = 0$$

When the *EM-GR* connection is in use these conditions will generate the splitting of the field equation analog to (4.2.2) into the standard *Compatibility Condition* and the second pair of Maxwell's equations. They become mandatory for maintaining compatibility with *General Relativity*. The resulting Lagrangian and field equations will be those from the *Four-Dimensional Conformal Gravitation* with extra torsion-dependent terms (that is, they vanish if the torsion does). By this way *Electromagnetism* becomes a direct manifestation of torsion and preserves its independence from *Gravitation*. Finally after adopting the splitting-values and renaming the constants by $(1,3,5,8) \rightarrow (1,2,3,4)$ the candidate action for unification results:

(5.1.13) *Conformal Electromagnetic Gravity Action*

$$I_{CEMG} \equiv k_0.\int_D \sqrt{-|g_{\bullet\bullet}|}.g^{kr}.g^{sl}.\Big( \overline{\overrightarrow{\tilde{R}}}_{kr}.\overline{\overrightarrow{\tilde{R}}}_{sl} + k_1.(\hat{\overrightarrow{\tilde{R}}}_{ks} - \tfrac{1}{2}.\overleftarrow{\tilde{R}}_{ks}).(\hat{\overrightarrow{\tilde{R}}}_{rl} - \tfrac{1}{2}.\overleftarrow{\tilde{R}}_{rl}).+$$
$$+ k_2.\partial\hat{\Gamma}_{ks}.\partial\hat{\Gamma}_{rl} + k_3.(\hat{\overrightarrow{\tilde{R}}}_{ks} - \tfrac{1}{2}.\overleftarrow{\tilde{R}}_{ks}).\partial\hat{\Gamma}_{rl} +$$
$$+ k_4.g^{mn}.g_{op}.(\tilde{R}^o{}_{ksm} + \tilde{R}^o{}_{mks} + \tilde{R}^o{}_{smk}).(\tilde{R}^p{}_{rln} + \tilde{R}^p{}_{nrl} + \tilde{R}^p{}_{lnr}) +$$
$$-(\tfrac{1}{2}.k_1 + \tfrac{3}{8}.k_3 + 3.k_4).(\tilde{R}^m{}_{kso} + \tilde{R}^m{}_{oks} + \tilde{R}^m{}_{sok}).(\tilde{R}^o{}_{rlm} + \tilde{R}^o{}_{mrl} + \tilde{R}^o{}_{lmr}) \Big).d\Omega$$

(5.1.14) / $$3.k_2 + k_3 \equiv \frac{32.k_g.\lambda_c}{3.c^4.c_1^2}$$

That the new terms are torsion-dependent can be seen by considering identities (2.2.10) and (2.2.22).

### 5.2 Field Equations

The corresponding field equations become:

$$
\begin{aligned}
(5.2.1)\quad \delta g^{ij} \;)\quad & \overline{\overrightarrow{\tilde{R}}}.(\overline{\overrightarrow{\tilde{R}}}_{ij}-\tfrac{1}{4}.\overline{\overrightarrow{\tilde{R}}}.g_{ij})= \\
&= k_1.(-(\hat{\overline{\overleftarrow{\tilde{R}}}}{}_i^{\;k}-\tfrac{1}{2}.\overleftarrow{\tilde{R}}{}_i^{\;k}).(\hat{\overline{\overleftarrow{\tilde{R}}}}_{jk}-\tfrac{1}{2}.\overleftarrow{\tilde{R}}_{jk})+\tfrac{1}{4}.(\hat{\overline{\overleftarrow{\tilde{R}}}}{}^{kr}-\tfrac{1}{2}.\overleftarrow{\tilde{R}}{}^{kr}).(\hat{\overline{\overleftarrow{\tilde{R}}}}_{kr}-\tfrac{1}{2}.\overleftarrow{\tilde{R}}_{kr}).g_{ij})+ \\
&\quad + k_2.(-\partial\hat{\Gamma}_i^{\;k}.\partial\hat{\Gamma}_{jk}+\tfrac{1}{4}.\partial\hat{\Gamma}^{kr}.\partial\hat{\Gamma}_{kr}.g_{ij})+ \\
&\quad + \tfrac{1}{2}.k_3.(-(\hat{\overline{\overleftarrow{\tilde{R}}}}{}_i^{\;k}-\tfrac{1}{2}.\overleftarrow{\tilde{R}}{}_i^{\;k}).\partial\hat{\Gamma}_{jk}-(\hat{\overline{\overleftarrow{\tilde{R}}}}{}_j^{\;k}-\tfrac{1}{2}.\overleftarrow{\tilde{R}}{}_j^{\;k}).\partial\hat{\Gamma}_{ik}+\tfrac{1}{2}.(\hat{\overline{\overleftarrow{\tilde{R}}}}{}^{kr}-\tfrac{1}{2}.\overleftarrow{\tilde{R}}{}^{kr}).\partial\hat{\Gamma}_{kr}.g_{ij})+ \\
&\quad + \tfrac{1}{2}.k_4.\Big(-3.(\tilde{R}_{pi}{}^{ln}+\tilde{R}_p{}^{n}{}_i{}^{l}+\tilde{R}_p{}^{ln}{}_i).(\tilde{R}^p{}_{jln}+\tilde{R}^p{}_{njl}+\tilde{R}^p{}_{lnj})+ \\
&\qquad + (\tilde{R}_i^{\;rln}+\tilde{R}_i^{\;nrl}+\tilde{R}_i^{\;lnr}).(\tilde{R}_{jrln}+\tilde{R}_{jnrl}+\tilde{R}_{jlnr})+ \\
&\qquad + \tfrac{1}{2}.(\tilde{R}_p{}^{rln}+\tilde{R}_p{}^{nrl}+\tilde{R}_p{}^{lnr}).(\tilde{R}^p{}_{rln}+\tilde{R}^p{}_{nrl}+\tilde{R}^p{}_{lnr}).g_{ij}\Big)+ \\
&\quad -(\tfrac{1}{2}.k_1+\tfrac{3}{8}.k_3+3.k_4).\Big(-(\tilde{R}^m{}_i{}^l{}_o+\tilde{R}^m{}_{oi}{}^l+\tilde{R}^{ml}{}_{oi}).(\tilde{R}^o{}_{jlm}+\tilde{R}^o{}_{mjl}+\tilde{R}^o{}_{lmj})+ \\
&\qquad + \tfrac{1}{4}.(\tilde{R}^{mrl}{}_o+\tilde{R}^m{}_o{}^{rl}+\tilde{R}^{ml}{}_o{}^r).(\tilde{R}^o{}_{rlm}+\tilde{R}^o{}_{mrl}+\tilde{R}^o{}_{lmr}).g_{ij}\Big)
\end{aligned}
$$

$$
\begin{aligned}
(5.2.2)\quad \delta\tilde{\Gamma}^i{}_{jk} \;)\quad & \tilde{\Pi}_i(\overline{\overrightarrow{\tilde{R}}}.g^{jk})-\delta_i^k.\tilde{\Pi}_r(\overline{\overrightarrow{\tilde{R}}}.g^{jr})+2.\hat{\Gamma}^k{}_{ir}.(\overline{\overrightarrow{\tilde{R}}}.g^{jr})= \\
&= -k_1.(\tilde{\Pi}_i(\hat{\overline{\overleftarrow{\tilde{R}}}}{}^{jk}-\tfrac{1}{2}.\overleftarrow{\tilde{R}}{}^{jk})-\delta_i^k.\tilde{\Pi}_r(\hat{\overline{\overleftarrow{\tilde{R}}}}{}^{jr}-\tfrac{1}{2}.\overleftarrow{\tilde{R}}{}^{jr})+2.\hat{\Gamma}^k{}_{ir}.(\hat{\overline{\overleftarrow{\tilde{R}}}}{}^{jr}-\tfrac{1}{2}.\overleftarrow{\tilde{R}}{}^{jr}))+ \\
&\quad -k_2.(\delta_i^k.(\tilde{\Pi}_p\partial\hat{\Gamma}^{pj}+\hat{\Gamma}^j{}_{pr}.\partial\hat{\Gamma}^{pr})-\delta_i^j.(\tilde{\Pi}_p\partial\hat{\Gamma}^{pk}+\hat{\Gamma}^k{}_{pr}.\partial\hat{\Gamma}^{pr}))+ \\
&\quad +k_1.\delta_i^j.(\tilde{\Pi}_p(\hat{\overline{\overleftarrow{\tilde{R}}}}{}^{pk}-\tfrac{1}{2}.\overleftarrow{\tilde{R}}{}^{pk})+\hat{\Gamma}^k{}_{rp}.(\hat{\overline{\overleftarrow{\tilde{R}}}}{}^{rp}-\tfrac{1}{2}.\overleftarrow{\tilde{R}}{}^{rp}))+ \\
&\quad -\tfrac{1}{2}.k_3.(\tilde{\Pi}_i\partial\hat{\Gamma}^{jk}-\delta_i^k.\tilde{\Pi}_p\partial\hat{\Gamma}^{jp}+2.\hat{\Gamma}^k{}_{ip}.\partial\hat{\Gamma}^{jp})+ \\
&\quad -\tfrac{1}{2}.k_3.\delta_i^k.(\tilde{\Pi}_p(\hat{\overline{\overleftarrow{\tilde{R}}}}{}^{pj}-\tfrac{1}{2}.\overleftarrow{\tilde{R}}{}^{pj})+\hat{\Gamma}^j{}_{pr}.(\hat{\overline{\overleftarrow{\tilde{R}}}}{}^{pr}-\tfrac{1}{2}.\overleftarrow{\tilde{R}}{}^{pr}))+ \\
&\quad +\tfrac{1}{2}.k_3.\delta_i^j.(\tilde{\Pi}_p(\hat{\overline{\overleftarrow{\tilde{R}}}}{}^{pk}-\tfrac{1}{2}.\overleftarrow{\tilde{R}}{}^{pk})+\hat{\Gamma}^k{}_{pr}.(\hat{\overline{\overleftarrow{\tilde{R}}}}{}^{pr}-\tfrac{1}{2}.\overleftarrow{\tilde{R}}{}^{pr}))+ \\
&\quad +\tfrac{1}{2}.k_3.\delta_i^j.(\tilde{\Pi}_p\partial\hat{\Gamma}^{pk}+\hat{\Gamma}^k{}_{rp}.\partial\hat{\Gamma}^{rp})+ \\
&\quad +2.(\tfrac{1}{2}.k_1+\tfrac{3}{8}.k_3+3.k_4).(\tilde{\Pi}_p(\tilde{R}^{kjp}{}_i+\tilde{R}^k{}_i{}^{jp}+\tilde{R}^{kp}{}_i{}^j)+\hat{\Gamma}^k{}_{rp}.(\tilde{R}^{pjr}{}_i+\tilde{R}^p{}_i{}^{jr}+\tilde{R}^{pr}{}_i{}^j)+ \\
&\qquad +\tilde{\Pi}_p(\tilde{R}^{jpk}{}_i+\tilde{R}^j{}_i{}^{pk}+\tilde{R}^{jk}{}_i{}^p)+\hat{\Gamma}^k{}_{rp}.(\tilde{R}^{jrp}{}_i+\tilde{R}^j{}_i{}^{rp}+\tilde{R}^{jp}{}_i{}^r)+ \\
&\qquad +\tilde{\Pi}_p(\tilde{R}^{pkj}{}_i+\tilde{R}^p{}_i{}^{kj}+\tilde{R}^{pj}{}_i{}^k)+\hat{\Gamma}^k{}_{rp}.(\tilde{R}^{rpj}{}_i+\tilde{R}^r{}_i{}^{pj}+\tilde{R}^{rj}{}_i{}^p))+ \\
&\quad -6.k_4.(\tilde{\Pi}_p(\tilde{R}_i^{\;jpk}+\tilde{R}_i^{\;kjp}+\tilde{R}_i^{\;pkj})+\hat{\Gamma}^k{}_{rp}.(\tilde{R}_i^{\;jrp}+\tilde{R}_i^{\;pjr}+\tilde{R}_i^{\;rpj}))
\end{aligned}
$$

This last equation can be shortly written as:

$$(5.2.3)\quad \tilde{\Pi}_i(\overline{\overrightarrow{\tilde{R}}}.g^{jk})-\delta_i^k.\tilde{\Pi}_r(\overline{\overrightarrow{\tilde{R}}}.g^{jr})+2.\hat{\Gamma}^k{}_{ir}.(\overline{\overrightarrow{\tilde{R}}}.g^{jr})=Q_i^{jk}$$

From this the following expression can be deduced:

$$(5.2.4)\quad \overline{\overrightarrow{\tilde{R}}}.\Big(g_{js}.(\tilde{\nabla}_i g^{sk}-g^{sk}.g_{ir}.\tilde{\nabla}_p g^{rp})+2.(\hat{\Gamma}^k{}_{ij}+\tfrac{1}{3}.(\delta_i^k.\hat{\Gamma}_j-\delta_j^k.\hat{\Gamma}_i))\Big)=Q_{ij}{}^k+\tfrac{1}{3}.(\delta_j^k.\bar{Q}_i-\delta_i^k.\bar{Q}_j)$$

$$(5.2.5)\quad / \qquad \bar{Q}^i \equiv Q_r{}^{ir}$$

If the generalized connection is expressed as the standard one plus as delta tensor the following identity holds:

(5.2.6) $$\tilde{\Gamma}^{i}{}_{jk} \equiv \Gamma^{i}{}_{jk} + \Delta^{i}{}_{jk} \quad \rightarrow \quad \tilde{\nabla}_{k} g^{ij} - g^{ij}.g_{kr}.\tilde{\nabla}_{s} g^{rs} = (\Delta^{ij}{}_{k} + \Delta^{ji}{}_{k}) - g^{ij}.g_{kr}.(\Delta^{rp}{}_{p} + \Delta^{pr}{}_{p})$$

This delta expression corresponds to the *Connection Compatibility Condition* (4.5.6).
Using this (5.2.4) translates into:

(5.2.7) $$\overline{\tilde{R}}.\left(\Delta_{j}{}^{k}{}_{i} + \Delta^{k}{}_{ji} - \delta^{k}_{j}.(\Delta_{i}{}^{p}{}_{p} + \Delta^{p}{}_{ip}) + 2.(\hat{\Delta}^{k}{}_{ij} + \tfrac{1}{3}.(\delta^{k}_{i}.\hat{\Delta}_{j} - \delta^{k}_{j}.\hat{\Delta}_{i}))\right) = Q_{ij}{}^{k} + \tfrac{1}{3}.(\delta^{k}_{j}.\vec{Q}_{i} - \delta^{k}_{i}.\vec{Q}_{j})$$

Starting with equation (5.2.3) by contraction of the index pair (ij) the following one can be derived:

(5.2.8) $$\overleftarrow{Q}^{i} \equiv Q_{r}{}^{ri} = 0 \quad \rightarrow \quad \tilde{\Pi}_{r}\partial\hat{\Gamma}^{ri} + \hat{\Gamma}^{i}{}_{sr}.\partial\hat{\Gamma}^{sr} = 0$$

By contraction of indexes (ik) on (5.2.3) this other is obtained:

(5.2.9) $$\partial_{i}(\ln \overline{\tilde{R}}) = -(g_{ir}.\tilde{\Pi}_{p} g^{rp} + \tfrac{2}{3}.\hat{\Gamma}_{i} + \tfrac{1}{3}.\overline{\tilde{R}}^{-1}.\vec{Q}_{i})$$

From this identity it can be seen that for adopting the *Relativistic Conformal Gauge* the following condition should be reached:

(5.2.10) $$\overline{\tilde{R}} \equiv 4.\lambda_{c} \quad \leftrightarrow \quad \int_{C} (12.\lambda_{c}.(g_{ir}.\tilde{\Pi}_{p} g^{rp} + \tfrac{2}{3}.\hat{\Gamma}_{i}) + \vec{Q}_{i}).dX^{i} = 0$$

According to (5.2.4) if the *Connection Compatibility Condition* (4.5.6) is met then the first equation can be decomposed into the following set which will be useful in later analysis:

(5.2.11) $$\tilde{\nabla}_{i} g^{sk} - g^{sk}.g_{ir}.\tilde{\nabla}_{p} g^{rp} \equiv 0$$ *Generalized Compatibility Condition*

(5.2.12) $$Q_{ij}{}^{k} = -Q_{ji}{}^{k}$$

(5.2.13) $$\rightarrow \quad 2.\overline{\tilde{R}}.(\hat{\Gamma}^{i}{}_{jk} + \hat{\Gamma}_{k}{}^{i}{}_{j} + \hat{\Gamma}_{jk}{}^{i}) = (Q_{jk}{}^{i} + Q^{i}{}_{jk} + Q_{k}{}^{i}{}_{j})$$

(5.2.14) $$2.\overline{\tilde{R}}.(\tfrac{1}{3}.(\delta^{i}_{j}.\hat{\Gamma}_{k} - \delta^{i}_{k}.\hat{\Gamma}_{j}) - \hat{\Gamma}_{k}{}^{i}{}_{j} - \hat{\Gamma}_{jk}{}^{i}) = \tfrac{1}{3}.(\delta^{i}_{j}.\vec{Q}_{k} - \delta^{i}_{k}.\vec{Q}_{j}) - Q_{k}{}^{i}{}_{j} - Q^{i}{}_{jk}$$

Considering (5.2.8), equality (5.2.12) implies the following one:

(5.2.15) $$\rightarrow \quad \vec{Q}^{i} = -\overleftarrow{Q}^{i} = 0$$

## 5.3 GR-Compatible Field Equations

When considering the *Electromagnetic Gravity Connection* the following identities hold:

(5.3.1) $$\tilde{\Gamma}^{i}{}_{jk} \equiv \Gamma^{i}{}_{jk} + \Theta^{i}{}_{jk} + c_{1}.\delta^{i}_{j}.A_{k}$$ *EM-GR Connection*

(5.3.2) $$\tilde{\nabla}_{i} g^{jk} - g^{jk}.g_{ir}.\tilde{\nabla}_{p} g^{rp} = 0$$

(5.3.3) $$\rightarrow \quad g_{ir}.\tilde{\Pi}_{p} g^{rp} + \tfrac{2}{3}.\hat{\Gamma}_{i} = 0 \quad , \quad \overleftarrow{Q}^{i} = 0$$

With connection (5.3.1) equation (5.2.8) transforms directly into the second pair of Maxwell's equations for null currents:

(5.3.4) (5.2.8) $\rightarrow \quad \nabla_k F^{jk} = 0$

All conditions (5.2.10), (5.2.11) and (5.2.12) are met:

(5.3.5) $\rightarrow \quad Q_{ij}^{\ \ k} = -Q_{ji}^{\ \ k} \quad , \quad Q_{jk}^{\ \ i} + Q^{i}_{\ jk} + Q_{k\ j}^{\ i} = 0 \quad , \quad \tfrac{1}{3}.(\delta^i_j.\bar{Q}_k - \delta^i_k.\bar{Q}_j) - Q_{k\ j}^{\ i} - Q^{i}_{\ jk} = 0$

The first pair of Maxwell's equations is obtained as an identity coming from the definition of the *Electromagnetic* field as the rotational of a vector potential, which was introduced with the *EM-GR Connection*:

(5.3.6) $F_{ij} \equiv \partial_i A_j - \partial_j A_i \quad \rightarrow \quad \partial_k F_{ij} + \partial_j F_{ki} + \partial_i F_{jk} = 0$

With identities seen on (5.3.3) equation (5.2.10) shows that for this connection the *Relativistic Conformal* gauge holds:

(5.3.7) (5.2.10) $\rightarrow \quad \overline{\vec{\tilde{R}}} = 4.\lambda_c$

Considering all of this and doing the corresponding substitutions the resulting compatible field equations become:

(5.3.8) $\Theta^{i}_{\ jk} \equiv \Theta^{i}_{\ kj} \quad , \quad \Theta_{i\ j}^{\ k} + \Theta^{k}_{\ ij} - \delta^k_i.(\Theta_{j\ r}^{\ r} + \Theta^{r}_{\ jr}) \equiv 0$

(5.3.9) $\nabla_i g^{jk} = 0$

(5.3.10) $\nabla_k F^{jk} = 0$

(5.3.11) $\nabla_k F_{ij} + \nabla_j F_{ki} + \nabla_i F_{jk} = 0$

(5.3.12) $R_{ij} - \tfrac{1}{2}.R.g_{ij} = \frac{8.\pi.k_g}{c^4}.T_{ij} - \lambda_c.g_{ij} + \frac{8.\pi.k_g}{c^4}.\left(\frac{1}{4.\pi}.(-F^k_i.F_{jk} + \tfrac{1}{4}.F^{kr}.F_{kr}.g_{ij})\right)$

(5.3.13) $T_{ij} \equiv \frac{c^4}{8.\pi.k_g}.\left(\overline{\vec{\Theta}}_{ij} - \tfrac{1}{2}.g^{kr}.\overline{\vec{\Theta}}_{kr}.g_{ij}\right)$

/

(5.3.14) $\overline{\vec{\Theta}}_{ij} \equiv -\nabla_k \Theta^{k}_{\ ij} + \Theta^{k}_{\ ir}.\Theta^{r}_{\ jk} + \tfrac{1}{2}.(\nabla_i \Theta^{k}_{\ jk} + \nabla_j \Theta^{k}_{\ ik}) - \Theta^{r}_{\ ij}.\Theta^{k}_{\ rk}$

The absence of current terms in (5.3.10) under the EM-GR connection is taken to be a consequence of a missing representation for particle fields. Such topic is a pending issue and requires a thinner analysis to be done in future research.

### 5.4 About the Current Term

For including the missing current in equation (5.3.10) one may be tempted to add to the Lagrangian a term of the form:

(5.4.1) $\sqrt{-|g_{\bullet\bullet}|}.g^{kr}.g^{sl}.k_5.J_{krs}.\hat{\Gamma}_l \quad / \quad J_{ijk} \equiv J_{jik}$

After selecting appropriately the corresponding $k_5$ constant the desired result would be obtained:

(5.4.2) $\nabla_k F^{jk} = -\frac{4.\pi}{c}.J^j \quad / \quad J_i \equiv J^{k}_{\ ki}$

Also a corresponding term would be added to the Einstein's stress-energy equation:

$$(5.4.3) \qquad R_{ij} - \tfrac{1}{2}.R.g_{ij} = ... + \tfrac{1}{8}.\lambda_c^{-1}.k_5.(-J_{ij}{}^k.\hat{\Gamma}_k - \tfrac{1}{2}.(J_i.\hat{\Gamma}_j + J_j.\hat{\Gamma}_i) + \tfrac{1}{2}.J^r.\hat{\Gamma}_r.g_{ij})$$

The term (5.4.1) can't be considered to be a constraining variable because that would force the nullity of the torsion's trace which in this context represents the *Electromagnetic* vector potential:

$$(5.4.4) \qquad \delta J_{ijk} \;) \qquad \hat{\Gamma}_i = 0 \qquad \rightarrow \qquad \text{Null } \textit{Electromagnetic} \text{ field ???}$$

Then currents should depend on the fundamental variables and demand new terms to be included in the current Lagrangian.

# 6 The Gauge Unification

## 6.1 Beyond Weyl's Theory

So far it was shown how the action proposed based on the *Geometrical Postulate* allows obtaining *Gravitation* and *Electromagnetism* together. Conformal invariance acting on the metric and lambda-invariance acting on the connection were introduced because the first imposes a four-dimensional manifold and permits recovering compatibility with *General Relativity* while the second allows to correctly introduce the *Electromagnetic* potential and field. The most interesting thing about having those single-object symmetries is this: they can be combined to represent the manifestation of a $U(1)$ gauge symmetry acting on a specific set of objects on the manifold (other than the metric and the connection) in a "geometric-like" transformation.

The effect on the action variables can be described by this transformation law:

$$(6.1.1) \qquad g'^{ij} \equiv e^{2.\varphi}.g^{ij} \qquad , \qquad \tilde{\Gamma}'^i{}_{jk} \equiv \tilde{\Gamma}^i{}_{jk} - \delta^i_j.\partial_k\varphi$$

This conjugation of symmetries does not eliminate any of them: is just a particular case but it will prove to be a useful one since it allows introducing the abelian gauge group needed. The transformation on the metric and the Lagrangian in use looks quite similar to those in Weyl's unification theory [8] (first and third terms of the Lagrange's scalar (5.1.13)) except that in this case a general connection with torsion is allowed and other terms related to *Electromagnetism* are considered. Also a different interpretation is given to the associated symmetry and the way it operates on tensors and it is shown on what conditions the compatibility with *General Relativity* can be recovered. This transformation won't be carried out on all tensors since some of them should remain invariant. For understanding this lets introduce the necessary concepts.

## 6.2 Diffeomorphisms and Parallel Transport

Consider a coordinate system change where each point $X$ is taken to a nearby one $X'$ by a linear transformation defined by a vector field $dX^i$:

$$(6.2.1) \qquad X'^i \equiv X^i + dX^i(X) \quad \rightarrow \quad \begin{cases} \dfrac{\partial X'^i}{\partial X^j} = \delta^i_j + \partial_j dX^i \\[2ex] \dfrac{\partial X^i}{\partial X'^j} \cong \delta^i_j - \partial_j dX^i \end{cases}$$

The representation for any covariant vector field $V^i$ in the new coordinate system can be obtained by applying the usual coordinate-transformation law:

$$(6.2.2) \qquad V'^i(X') = \frac{\partial X'^i}{\partial X^k}.V^k(X) = V^i + V^k.\partial_k dX^i$$

This expression relates representations of the same vector field evaluated at different points with different coordinate systems. It is not an identity between tensors evaluated at the same point.
I will become "*tensorial*" once the first operand is expressed in the initial coordinate system:

(6.2.3) $$V'^i(X^i + dX^i) = V'^i(X) + dX^k(X).\partial_k V'^i(X) + ...$$

(6.2.4) $$\rightarrow \qquad V'^i \cong V^i - \underset{(dX)}{L} V^i$$

(6.2.5) $$/ \qquad \underset{(dX)}{L} V^i \equiv dX^k.\partial_k V^i - V^k.\partial_k dX^i \qquad \textit{Lie's derivative along } dX^i$$

Expression (6.2.4) is equivalent to (6.2.2) and have all the fields evaluated on the initial coordinate system. The transformation's operator results to be the Lie derivative.

The diffeomorphism considered corresponds to a change of coordinates where the tensor fields are not being modified. In that sense it's a "*passive*" transformation.
Now consider the following "*active*" one: take a vector field $V^i$ and parallel-transport it using the manifold's generic connection $\widetilde{\Gamma}^k{}_{ij}$ along a given displacement field $dX^i$:

(6.2.6) $$V'^i(X + dX) = V^i(X) + dX^r(X).\widetilde{\nabla}_k V^i(X)$$

For extracting a tensorial equation out of this expression a coordinate change is done so that the starting point $X$ is taken to the ending one $(X + dX)$ along $dX^i$:

(6.2.7) $$V'^i(X) \cong V^i + (\widetilde{\nabla}_k dX^i - 2.\hat{\Gamma}^i{}_{rk}.dX^r).V^k$$

This active transformation up to a first differential order becomes characterized by the product of the vector field with the following tensor operator:

(6.2.8) $$V'^i \cong U^i{}_k.V^k \qquad / \qquad U^i{}_j \equiv \delta^i_j + \widetilde{\nabla}_j dX^i - 2.\hat{\Gamma}^i{}_{rj}.dX^r = \delta^i_j + \overline{\nabla}_j dX^i - \hat{\Gamma}^i{}_{rj}.dX^r$$

The resulting linear operator $U^i{}_j$ only contains elements associated to the transformation itself and will be the same no matter the transforming vector considered. It's interesting to notice that this operator depends on a connection having an opposite torsion with respect to the original one.

## 6.3 Loop Transformations

If the parallel transport of a vector field is done along a closed infinitesimal curve $C$ the ending point becomes the original one so there is no need to apply any coordinate system change since the equation obtained results tensorial. In that case the transformation will be called *Loop Transformation*.
Such change will depend on the curve considered and translates into a tensor operator involving the general connection and tensors derived from it. The simplest case is obtained when the curve is a "parallelogram" defined by two infinitesimal vector fields $dX^i$ and $dY^i$ (simplest kind of loop) which were parallel transported along each other for obtaining the opposite sides and the gap caused by the torsion was appended for having the necessary closure.

The resulting operator can then be expressed using the corresponding curvature tensor:

(6.3.1) $$V'^i \cong U^i{}_k.V^k \qquad / \qquad U^i{}_j \equiv \delta^i_j + \widetilde{R}^i{}_{jrs}.dX^r dY^s$$

### 6.4 Absolute and Relative Transformations

Looking to the previous transformations two kinds can be distinguished: those that depend on the field being transformed and those do not. The firsts are linear operators containing derivatives to be evaluated on the transformed field. Here they will be called *Relative Transformations*. The seconds become tensor fields depending only on the transformation agents and will be referred as *Absolute Transformations*.

Among this terminology the first example (6.2.4) becomes a relative transformation while the other two (6.2.8) and (6.3.1) result to be absolute.

### 6.5 Absolute Transformations as a Group

Absolute transformations are the most interesting entities since they exist as independent stand-alone fields. They interact with all transformed fields in a homogeneous way and can be generally represented as the following sum:

(6.5.1) $$U^i{}_j \equiv \delta^i_j + \underset{(a)}{T^i{}_j}.\underset{(a)}{d\theta} + \underset{(ab)}{T^i{}_j}.\underset{(a)}{d\theta}.\underset{(b)}{d\theta} + \underset{(abc)}{T^i{}_j}.\underset{(a)}{d\theta}.\underset{(b)}{d\theta}.\underset{(c)}{d\theta} + \ldots$$

Where $\underset{(a)}{T^i{}_j}, \underset{(ab)}{T^i{}_j}, .., \underset{(a..z)}{T^i{}_j}$ are arbitrary tensors which define the transformation and $\underset{(a)}{d\theta}$ a finite set of infinitesimal scalar-functions acting as driving parameters.
These transformations are invertible since an inverse can always be found:

(6.5.2) $$\exists\ \overset{-1}{U^i{}_j}\ \ /\ \ U^i{}_k.\overset{-1}{U^k{}_j} = \delta^i_j$$

(6.5.3) $$\overset{-1}{U^i{}_j} \equiv \delta^i_j + \underset{(a)}{V^i{}_j}.\underset{(a)}{d\theta} + \underset{(ab)}{V^i{}_j}.\underset{(a)}{d\theta}.\underset{(b)}{d\theta} + \underset{(abc)}{V^i{}_j}.\underset{(a)}{d\theta}.\underset{(b)}{d\theta}.\underset{(c)}{d\theta} + \ldots$$

$$/\ \left\{\begin{array}{ll} \underset{(a)}{V^i{}_j} = -\underset{(a)}{T^i{}_j} & (6.5.4)\\ \underset{(ab)}{V^i{}_j} = -\underset{(ab)}{T^i{}_j} + \underset{(a)}{T^i{}_k}.\underset{(b)}{T^k{}_j} & (6.5.5)\\ \underset{(abc)}{V^i{}_j} = -\underset{(abc)}{T^i{}_j} + \underset{(a)}{T^i{}_k}.\underset{(bc)}{T^k{}_j} + \underset{(ab)}{T^i{}_k}.\underset{(c)}{T^k{}_j} - \underset{(a)}{T^i{}_k}.\underset{(b)}{T^k{}_r}.\underset{(c)}{T^r{}_j} & (6.5.6)\\ \ldots & \end{array}\right.$$

They represent a transformation group which will be called *Absolute Transformation Group*.

### 6.6 Absolute Transformations as the Key Invariance Symmetry

In the plan of constructing a field theory based on the manifold's geometric structure only absolute transformations based on geometrical objects will be considered. As shown in the previous examples they can be defined by combining parallel displacements and loop transformations which are well understood geometrical operations but in principle any transformation build out of geometrical objects will be good even if it doesn't have a direct interpretation as a composition of known "moves". Of course knot-based loops will generate absolute loop transformations and in that case the resulting transformations may be classified according to the corresponding loop-invariants.

Invariance under the action of absolute transformations will be the key principle of the theory.
For being generic enough the geometrical objects considered will be displacement fields, the metric, the general connection and gauge group generators which will be introduced later:

(6.6.1)

*AT-Geometrical Field Theory* $\leftrightarrow$ *Field equations are covariant under* $AT(dX^{\bullet}, g_{\bullet\bullet}, \tilde{\Gamma}^{\bullet}{}_{\bullet\bullet}, \underset{(x)}{E^{\bullet}{}_{\bullet}})$.

## 6.7 Invariant Tensors

During absolute transformations some tensors remain unchanged.
The manifold's points remain invariant under an absolute loop transformation (*ALT*) so any displacement vector field linking two points will remain the same:

(6.7.1) $$X'^i = X^i \quad \rightarrow \quad dX'^i(X) = dX^i(X) \qquad \text{(ALT)}$$

Any scalar field also remains the same under an *ALT* and so do the contraction between a covariant and a contravariant vector fields. If both are transforming fields this implies that the contravariant transforming tensor should be the inverse of the covariant one:

(6.7.2) $$\left.\begin{array}{l} V'^i = U^i{}_k.V^k \\ W'_i = \widehat{U}^k{}_i.W_k \end{array}\right\} \rightarrow \phi'(X) = \phi(X) \rightarrow V'^k.W'_k = V^k.W_k \ \ldots \ \ldots \rightarrow \widehat{U}^i{}_k.U^k{}_j = \delta^i_j \rightarrow \widehat{U}^i{}_j = \overset{-1}{U^i{}_j}$$

The *Kronecker-delta* or *Identity* tensor remains invariant under any absolute transformation (*AT)* since covariant and contravariant transforming tensors are inverses from one another:

(6.7.3) $$\delta'^i_j = U^i{}_k.\delta^k_r.\overset{-1}{U^r{}_j} = \delta^i_j \qquad \text{(AL)}$$

From (6.7.1) and (6.7.2) results that the gradient of a scalar function is also invariant under any *ALT*:

(6.7.4) $$\left.\begin{array}{ll} d\phi'(X) = d\phi(X) & \text{(ALT)} \\ dX'^i(X) = dX^i(X) & \text{(ALT)} \\ d\phi = \partial_k\phi.dX^k & \end{array}\right\} \rightarrow \partial_i\phi'(X) = \partial_i\phi(X) \qquad \text{(ALT)}$$

It will be seen that gauge fields will be included into the theory by imposing its Lagrangian to be absolute-loop invariant under suitable transformations based on the corresponding gauge group generators.

## 6.8 Index Rules and Types

For being consistent with a different transformation law for transforming and invariant vector fields, only vectors (indexes) of the same kind can be added or contracted. The sum or contraction of different kind of vectors results in a tensor having an arbitrary transformation law were the absolute characteristic of the transformation is lost:

(6.8.1) $$\left.\begin{array}{l} V'^i = U^i{}_k.V^k \\ W'^i = W^k \\ Y^i \equiv V^i + W^i \ ? \end{array}\right\} \rightarrow \left\{\begin{array}{l} Y'^i = U^i{}_k.V^k + W^k \neq U^i{}_k.(V^k + W^k) \neq V^i + W^i \\ V'^k.W'_k = U^k{}_r.V^r.W_k \neq V^k.W_k \end{array}\right.$$

Exceptions to this rule apply to tensors that can be considered to be either transforming or invariant. A general case would be those proportional to the identity tensor. A particular case for a given transforming operator $U^i{}_j$ would be those tensors commuting with it:

(6.8.2) $$\left.\begin{array}{l} A'^i{}_j = U^i{}_k.A^k{}_r.U^{-1^r}{}_j \\ B'^i{}_j = B^i{}_j \\ C^i{}_j \equiv A^i{}_j + \phi.\delta^i_j \ ? \\ D^i{}_j \equiv B^i{}_j + \phi.\delta^i_j \ ? \end{array}\right\} \rightarrow \left\{\begin{array}{l} C'^i{}_j = U^i{}_k.(A^k{}_r + \phi.\delta^k_r).U^{-1^r}{}_j \\ D'^i{}_j = B^i{}_j + \phi.\delta^i_j \end{array}\right.$$

(6.8.3) $$\left.\begin{array}{l} A'^i{}_j = U^i{}_k.A^k{}_r.U^{-1}{}^r{}_j \\ U^i{}_k.A^k{}_j - A^i{}_k.U^k{}_j \equiv 0 \end{array}\right\} \quad \rightarrow \quad A'^i{}_j = U^i{}_k.A^k{}_r.U^{-1}{}^r{}_j = A^i{}_k.U^k{}_r.U^{-1}{}^r{}_j = A^i{}_j$$

The transforming tensors $U^i{}_j$ when acting on themselves can also be considered as transforming or invariant.

Self-contraction is allowed whenever contracted indexes are of the same type:

(6.8.4) $$A'^k{}_k = U^k{}_s.A^s{}_r.U^{-1}{}^r{}_k = A^k{}_k$$

To resume there are three kinds of indexes: the *Transforming* ones, the *Invariant* ones and those like in the identity tensor which act in pairs of mixed indexes (covariant and contravariant) that can be considered to be either transforming or invariant. These last will be called *Neutral* indexes.

### 6.9 Connection Index Types

Consider the covariant differential on any vector field:

(6.9.1) $$\widetilde{D}V^i \equiv \widetilde{\nabla}_r V^i.dX^r = (\partial_r V^i + \widetilde{\Gamma}^i{}_{kr}.V^k).dX^r$$

(6.9.2) $$\widetilde{D}W_i \equiv \widetilde{\nabla}_r W_i.dX^r = (\partial_r W_i - \widetilde{\Gamma}^k{}_{ir}.W_k).dX^r$$

The connection's last covariant index is contracted with $dX^i$ being an *ALT*-invariant vector so it should be also *ALT*-invariant according to index rules. The other connection indexes may be contracted with either transforming or invariant vectors since the same connection is used with all of them so they should behave as a *neutral-pair*.

(6.9.3) $$\widetilde{\Gamma}^i{}_{jk} \quad \rightarrow \quad k\text{- Invariant} \quad , \quad i, j - \text{Neutral}$$

### 6.10 The Connection transformation law

The transformation law for the connection can be obtained by demanding absolute transformations to commute with the covariant derivative operation:

(6.10.1) $$\widetilde{\nabla}'_j V'^i \equiv (\widetilde{\nabla}_j V^i)'$$

There are two possible ways to define such relation depending on the type of vector considered.
For transforming vector fields this yields the following law:

(6.10.2) $$V'^i = U^i{}_k.V^k \quad \rightarrow \quad \widetilde{\Gamma}'^i{}_{jk} = \overset{-1}{U}{}_j{}^r.(U^i{}_l.\widetilde{\Gamma}^l{}_{rk} - \partial_k U^i{}_r) = \widetilde{\Gamma}^i{}_{jk} + U^i{}_r.\widetilde{\nabla}_k \overset{-1}{U}{}^r{}_j$$

For invariant vector fields the law should read:

(6.10.3) $$W'^i = W^k \quad \rightarrow \quad \widetilde{\Gamma}'^i{}_{jk} = \widetilde{\Gamma}^i{}_{jk}$$

These two transformation laws can coexist if the theory fixes the connection field up to a degree of freedom given by the extension $U^i{}_r.\widetilde{\nabla}_k \overset{-1}{U}{}^r{}_j$ produced by the absolute transformation in play.
In other words connections differing only on an *AT*-extension should be regarded as representing the same geometrical object which becomes an equivalence class under these *AT*.

The theory's field equations should only be sensible to the connection's equivalent class and completely invariant to the action of the proposed transformations. So do the resulting fields and all the physics involved. As a consequence of this the connection can be considered to follow transformation law (6.10.2) and all physics equations should be invariant under these changes.

When considering a transformation up to a first order in the transforming parameters, the connection will change as follows:

$$(6.10.4)\quad \underset{(a)}{U^i}_j = \delta^i_j + \underset{(a)}{T^i}_j . \underset{(a)}{d\theta} + \dots \;, \qquad \overset{-1}{\underset{(a)}{U^i}}_j = \delta^i_j - \underset{(a)}{T^i}_j . \underset{(a)}{d\theta} + \dots$$

$$(6.10.5)\quad \rightarrow \qquad \tilde{\Gamma}'^i{}_{jk} \cong \tilde{\Gamma}^i{}_{jk} - \tilde{\nabla}_k(\underset{(a)}{T^i}_j . \underset{(a)}{d\theta})$$

## 6.11 Conformal Electromagnetic Gravitation as an Absolute Invariant theory

Conformal and lambda transformations can be conjugated to represent and absolute-loop transformation corresponding to a $U(1)$ gauge group, the one associated to *Electromagnetism* (although particle wave functions and phase invariance were not introduced). The corresponding operators and the effects on the connection are:

$$(6.11.1)\quad \underset{U(1)}{U^i}_j \equiv \delta^i_j . \mathrm{e}^{c_1.\varphi} \qquad , \qquad \underset{U(1)}{U^i}_j{}^{-1} \equiv \delta^i_j . \mathrm{e}^{-c_1.\varphi}$$

$$(6.11.2)\quad \rightarrow \qquad \tilde{\Gamma}'^i{}_{jk} = \tilde{\Gamma}^i{}_{jk} + U^i{}_r . \tilde{\nabla}_k \overset{-1}{U^r}_j = \tilde{\Gamma}^i{}_{jk} - c_1 . \delta^i_j . \partial_k \varphi$$

The metric will be considered to be a transforming tensor:

$$(6.11.3)\quad g'^{ij} \equiv \underset{U(1)}{U^i}_k . \underset{U(1)}{U^i}_r . g^{kr} = \mathrm{e}^{2.c_1.\varphi} . g^{ij}$$

Considering the Lagrangian terms:

- The metric is an absolute-transforming tensor but it changes conformally so it does not affect the current Lagrangian nor the physics involved.
- The curvature results invariant since the absolute-transformation produces a lambda one on the connection.
- The torsion-trace's rotational terms are also invariant for the lambda transformation.

Because of these the Lagrangian results invariant and the *Conformal Electromagnetic Gravitation* becomes an absolute-invariant theory with respect to transformations (6.11.1).

The point of having a conformal symmetry in the theory now becomes clear: the *Electromagnetic* gauge effect on the metric can be canceled out by a suitable conformal transformation so the metric and the associated geometry may stand still for any observer.

The $U(1)$ gauge freedom will contribute to the *Global Phase* when considering the *Relativistic Conformal Gauge*:

$$(6.11.4)\quad \tilde{\phi}(C,X) \equiv -\int_C (g_{sr}.\tilde{\Pi}_k g^{rk} + \tfrac{2}{3}.\hat{\Gamma}_s).dX^s - 2.c_1 . \underset{(E.M.)}{\phi} = -\int_C (g_{sr}.\tilde{\Pi}_k g^{rk} - c_1.(A_s - 2.\partial_s \underset{(E.M.)}{\phi})).dX^s$$

In that case the metric-conformal gauge will be also compensating the *Electromagnetic* gauge when reaching compatibility.

### 6.12 The Electromagnetic Potential's transformation law: is the Higgs boson necessary?

Considering the *EM-GR* connection, the transformation law for a $U(1)$ regauging becomes:

(6.12.1) $$\widetilde{\Gamma}^{i}{}_{jk} = \Gamma^{i}{}_{jk} + \Theta^{i}{}_{jk} + c_1.\delta^{i}_{j}.A_k \quad \rightarrow \quad \widetilde{\Gamma}'^{i}{}_{jk} = \Gamma^{i}{}_{jk} + \Theta^{i}{}_{jk} + c_1.\delta^{i}_{j}.(A_k - \partial_k \varphi)$$

It looks like the potential is transforming according to the standard gauge rule for *Electromagnetism*:

(6.12.2) $$A_k \quad \rightarrow \quad A'_k = A_k - \partial_k \varphi$$

But due to absolute-invariance index behavior the vector potential comes to be invariant making the previous transformation rule not a true one. The good one should be:

(6.12.3) $$A_k \quad \rightarrow \quad A'_k = A_k$$

The gauge extension-term appearing in the connection is caused by the connection transformation law not by the vector potential one. This enables the vector potential to be placed directly into the Lagrangian. The corresponding contributions to the field equations become gauge invariant. This fact turns out to be quite interesting since it allows in particle-theory formulation to introduce mass-terms based on the field's potential in order to give mass to the corresponding boson fields. The actual Higgs mechanism used for that purpose in the *Standard Model* [7] may not be necessary at all since mass terms can be introduced without breaking any symmetry. That may happen if particle physics and the *Standard Model* can be both reformulated while being based on the present theory.

(6.12.4) *Standard Model's Lagrangian* $\rightarrow \quad ... + g^{kr}.g^{sl}.m_{sl}.A_k.A_r + ...$ *Mass terms are allowed*

# 7 Introducing Other Gauge Symmetries

## 7.1 Gauge Field components

Following Yang-Mills steps once *Electromagnetism* has been addressed the next step is to generalize the theory for allowing other gauge groups. Since the *Conformal Electromagnetic Gravitation* works in a four-dimensional manifold it turns out that not all the symmetries of the *Standard Model* can be introduced while using a real domain just because there are not enough dimensions. At least there is space for introducing another $SU(2)\times SU(2)$ representation without having to talk about an extra *Internal Space*.

For introducing the components let's begin with a generic gauge group.
The elements defining the group are:

(7.1.1) $\underset{(a)}{E^{i}{}_{j}}(X)$ *Group Generators* / $$\underset{(a)}{E^{i}{}_{k}}.\underset{(b)}{E^{k}{}_{j}} - \underset{(b)}{E^{i}{}_{k}}.\underset{(a)}{E^{k}{}_{j}} \equiv \underset{(cab)}{C}.\underset{(c)}{E^{i}{}_{j}}$$

(7.1.2) $\underset{(cab)}{C}$ *Structural Constants* / $$\begin{cases} \underset{(cab)}{C} \equiv -\underset{(cba)}{C} \\ \underset{(eda)}{C}.\underset{(dbc)}{C} + \underset{(edc)}{C}.\underset{(dab)}{C} + \underset{(edb)}{C}.\underset{(dca)}{C} \equiv 0 \end{cases}$$

(7.1.3) $\underset{(a)}{A_j}(X)$ *Vector Potentials*

(7.1.4) $\underset{(a)}{H_{ij}}(X)$ *Gauge Fields* / $$\underset{(a)}{H_{ij}} \equiv \partial_i \underset{(a)}{A_j} - \partial_j \underset{(a)}{A_i} + c_3.\underset{(abc)}{C}.\underset{(b)}{A_i}.\underset{(c)}{A_j}$$

Where $\underset{(a)}{E^i{}_j}$ are tensor fields having the role of group generators, $\underset{(cab)}{C}$ are the group's structural constants, $\underset{(a)}{A_j}$ the field potentials, $c_3$ their corresponding coupling constant and $\underset{(a)}{H_{ij}}$ the associated gauge fields. They will be integrated into the generalized connection in the following way:

$$\widetilde{\Gamma}^i{}_{jk} \equiv \Gamma^i{}_{jk} + \Theta^i{}_{jk} + c_1.\delta^i_j.A_k - c_3.\underset{(a)}{E^i{}_j}.\underset{(a)}{A_k} \tag{7.1.5}$$

Like in Yank-Mills theory the transforming operator will be:

$$U^i{}_j \equiv Exp(c_3.\underset{(a)}{E^i{}_j}(X).\underset{(a)}{\varphi}(X)) \quad , \qquad \overset{-1}{U^i{}_j} \equiv Exp(-c_3.\underset{(a)}{E^i{}_j}(X).\underset{(a)}{\varphi}(X)) \tag{7.1.6}$$

The generator's trace should vanish causing the operator's determinant to have a unit value:

$$\underset{(a)}{E^k{}_k} \equiv 0 \qquad \rightarrow \qquad \det\left|\underset{(a)}{U^i{}_j}\right| = 1 \tag{7.1.7}$$

Properties (7.1.1) and (7.1.7) are absolute-invariant with respect to *Electromagnetism* (6.11.1) and the actual *Gauge* transformations (7.1.6) so they can be imposed using invariant constraining fields $(\widetilde{C}' = \widetilde{C})$.

As seen in (6.10.5), on any infinitesimal transformation defined by $\underset{(a)}{\delta\varphi}$ the connection transforms like:

$$\widetilde{\Gamma}'^i{}_{jk} \cong \widetilde{\Gamma}^i{}_{jk} - c_3.(\widetilde{\nabla}_k \underset{(a)}{E^i{}_j}.\underset{(a)}{\delta\varphi} + \underset{(a)}{E^i{}_j}.\partial_k \underset{(a)}{\delta\varphi}) \tag{7.1.8}$$

This *ALT*-extension will not leave the Lagrangian invariant unless the group representation and the transforming parameters fulfill some extra conditions.

## 7.2 Metric Constraint and the Homogenous Lorentz Group

The metric was taken to be a transforming tensor when applying the $U(1)$ absolute transformation associated to *Electromagnetism* and compatibility with *General Relativity* was recovered by selecting the *Relativistic Conformal Gauge*. For any other absolute transformation even if compatibility is not met the Lagrangian should remain invariant. This will be achieved by granting the invariance of its components like the curvature, the torsion-trace's rotational and the metric. Since there is no other symmetry like the conformal one that can be used for compensating changes on this last tensor the allowed transformations should be constrained to those of the *Homogenous Lorentz Group* which are the ones leaving the metric invariant. From this the Lorentz constraint results:

$$g'_{ij} = U^k{}_i.U^r{}_j.g_{kr} \equiv g_{ij} \qquad \textit{Lorentz Constraint} \tag{7.2.1}$$

This can be traduced into symmetries on the group generator's indexes. When considering an infinitesimal transformation up to a first order this definition turns out to be:

$$\underset{(a)}{E_{ij}} + \underset{(a)}{E_{ji}} \equiv 0 \quad / \quad a \in \{1,2,3,4,5,6\} \qquad \textit{Group Generator's Symmetry} \tag{7.2.2}$$

The group generators are six skew-symmetric tensors in their covariant form and their traces become null as expressed earlier in (7.1.7).

Conditions (7.2.2) are *AT*-invariant with respect to *Electromagnetism* and *Gauge* transformations so they can also be imposed using an invariant constraining field.

From an inertial coordinate system these tensors can be taken to be the following set:

*Lorentz Group Generators*

(7.2.3)
$$\left\{\begin{array}{lll}
\underset{(1)}{E_{ij}} \equiv \begin{bmatrix} & -1 & & \\ +1 & & & \\ & & & \\ & & & \end{bmatrix} & \underset{(2)}{E_{ij}} \equiv \begin{bmatrix} & & -1 & \\ & & & \\ +1 & & & \\ & & & \end{bmatrix} & \underset{(3)}{E_{ij}} \equiv \begin{bmatrix} & & & -1 \\ & & & \\ & & & \\ +1 & & & \end{bmatrix} \\
\underset{(4)}{E_{ij}} \equiv \begin{bmatrix} & & & \\ & & & \\ & & & +1 \\ & & -1 & \end{bmatrix} & \underset{(5)}{E_{ij}} \equiv \begin{bmatrix} & & & \\ & & & -1 \\ & & & \\ & +1 & & \end{bmatrix} & \underset{(6)}{E_{ij}} \equiv \begin{bmatrix} & & & \\ & & +1 & \\ & -1 & & \\ & & & \end{bmatrix}
\end{array}\right.$$

After raising the first index with the metric they become:

(7.2.4)
$$g^{ij} \equiv \begin{bmatrix} +1 & & & \\ & -1 & & \\ & & -1 & \\ & & & -1 \end{bmatrix} \quad \rightarrow \quad \underset{(a)}{E^{i}{}_{j}} = g^{ik}.\underset{(a)}{E_{kj}}$$

$$\rightarrow \left\{\begin{array}{lll}
\underset{(1)}{E^{i}{}_{j}} \equiv \begin{bmatrix} & -1 & & \\ -1 & & & \\ & & & \\ & & & \end{bmatrix} & \underset{(2)}{E^{i}{}_{j}} \equiv \begin{bmatrix} & & -1 & \\ & & & \\ -1 & & & \\ & & & \end{bmatrix} & \underset{(3)}{E^{i}{}_{j}} \equiv \begin{bmatrix} & & & -1 \\ & & & \\ & & & \\ -1 & & & \end{bmatrix} \\
\underset{(4)}{E^{i}{}_{j}} \equiv \begin{bmatrix} & & & \\ & & & \\ & & & -1 \\ & & +1 & \end{bmatrix} & \underset{(5)}{E^{i}{}_{j}} \equiv \begin{bmatrix} & & & \\ & & & +1 \\ & & & \\ & -1 & & \end{bmatrix} & \underset{(6)}{E^{i}{}_{j}} \equiv \begin{bmatrix} & & & \\ & & -1 & \\ & +1 & & \\ & & & \end{bmatrix}
\end{array}\right.$$

These tensors can be arranged in two sets: one corresponding to space-time accelerations $\underset{(x)}{K^{i}{}_{j}}$ and the other associated to space rotations $\underset{(x)}{L^{i}{}_{j}}$:

(7.2.5)
$$\left\{\begin{array}{llll}
\underset{(x)}{K^{i}{}_{j}} \equiv \underset{(1)}{E^{i}{}_{j}} , & \underset{(y)}{K^{i}{}_{j}} \equiv \underset{(2)}{E^{i}{}_{j}} , & \underset{(z)}{K^{i}{}_{j}} \equiv \underset{(3)}{E^{i}{}_{j}} & \textit{Acceleration} \\
\underset{(x)}{L^{i}{}_{j}} \equiv \underset{(4)}{E^{i}{}_{j}} , & \underset{(y)}{L^{i}{}_{j}} \equiv \underset{(5)}{E^{i}{}_{j}} , & \underset{(z)}{L^{i}{}_{j}} \equiv \underset{(6)}{E^{i}{}_{j}} & \textit{Rotation}
\end{array}\right.$$

Combining these tensors the following complex set of group generators can be formed:

(7.2.6)
$$\underset{(a)}{X^{\pm i}{}_{j}} \equiv \tfrac{1}{2}.(\underset{(a)}{K^{i}{}_{j}} \pm i.\underset{(a)}{L^{i}{}_{j}}) \quad / \quad a \in \{x,y,z\} \qquad \textit{Lorentz Complex Group Generators}$$

(7.2.7)
$$\underset{(x)}{X^{+i}{}_{j}} \equiv \tfrac{1}{2}\begin{bmatrix} & -1 & & \\ -1 & & & \\ & & & -i \\ & & +i & \end{bmatrix} \quad \underset{(y)}{X^{+i}{}_{j}} \equiv \tfrac{1}{2}\begin{bmatrix} & & -1 & \\ & & & +i \\ -1 & & & \\ & -i & & \end{bmatrix} \quad \underset{(z)}{X^{+i}{}_{j}} \equiv \tfrac{1}{2}\begin{bmatrix} & & & -1 \\ & & -i & \\ & +i & & \\ -1 & & & \end{bmatrix}$$

(7.2.8)
$$\underset{(x)}{X^{-i}{}_{j}} \equiv \tfrac{1}{2}\begin{bmatrix} & -1 & & \\ -1 & & & \\ & & & +i \\ & & -i & \end{bmatrix} \quad \underset{(y)}{X^{-i}{}_{j}} \equiv \tfrac{1}{2}\begin{bmatrix} & & -1 & \\ & & & -i \\ -1 & & & \\ & +i & & \end{bmatrix} \quad \underset{(z)}{X^{-i}{}_{j}} \equiv \tfrac{1}{2}\begin{bmatrix} & & & -1 \\ & & +i & \\ & -i & & \\ -1 & & & \end{bmatrix}$$

These generators splits the *Lorentz Group* into the direct sum of two $SU(2)$ algebras as it can be seen in their commutation rules:

(7.2.9)
$$\left[\underset{(a)}{X^{+}},\underset{(b)}{X^{+}}\right] = i.\underset{(abc)}{\varepsilon}.\underset{(c)}{X^{+}}$$

(7.2.10)
$$\left[\underset{(a)}{X^{-}},\underset{(b)}{X^{-}}\right] = -i.\underset{(abc)}{\varepsilon}.\underset{(c)}{X^{-}}$$

(7.2.11) $$\left[ \underset{(a)}{X^+}, \underset{(b)}{X^-} \right] = 0$$

$$/ \qquad a,b,c \in \{x,y,z\}$$

The algebra (7.2.10) corresponds to a $SU(2)$ one for a reversed orientation on $\{x,y,z\}$.
They also fulfill the following properties (no contraction is being carried on the space index $(x)$):

(7.2.12) $$\underset{(x)}{X^{+}}{}^i{}_k . \underset{(x)}{X^{+}}{}^k{}_j = \underset{(y)}{X^{+}}{}^i{}_k . \underset{(y)}{X^{+}}{}^k{}_j = \underset{(z)}{X^{+}}{}^i{}_k . \underset{(z)}{X^{+}}{}^k{}_j = \underset{(x)}{X^{-}}{}^i{}_k . \underset{(x)}{X^{-}}{}^k{}_j = \underset{(y)}{X^{-}}{}^i{}_k . \underset{(y)}{X^{-}}{}^k{}_j = \underset{(z)}{X^{-}}{}^i{}_k . \underset{(z)}{X^{-}}{}^k{}_j = \tfrac{1}{4}.\delta^i_j$$

(7.2.13) $$\underset{(a)}{X^{\pm}}{}^k{}_k = 0$$

The *Lorentz* group representation is isomorphic to the algebra $SU(2)\times SU(2)$ so the Lie-Algebra structure of the theory can be taken to be:

(7.2.14) *Theory's Lie-Algebra* $$\rightarrow \qquad GR \times \underset{EM}{U(1)} \times \underset{L^+}{SU(2)} \times \underset{L^-}{SU(2)}$$

Where $GR$ symbolize the *Gravity* contribution, $\underset{EM}{U(1)}$ the *Electromagnetic* one and the last two correspond to the *Lorentz* group in its complex representation.

## 7.3 Group Generators Constraint

Other important terms in the Lagrangian are those related to the curvature tensor. For any infinitesimal absolute transformation (7.1.6) this will remain invariant up to a first order if the following differential conditions hold for the group generators:

(7.3.1) $$\left. \begin{array}{l} \widetilde{\nabla}_k \underset{(a)}{E^i}{}_j \equiv 0 \\ U^i{}_j \cong \delta^i_j + \varepsilon . c_3 . \underset{(a)}{E^i}{}_j . \underset{(a)}{\varphi} \\ \quad / \quad \varepsilon << 1 \end{array} \right\} \rightarrow \quad \widetilde{R}'^i{}_{jkl} = \widetilde{R}^i{}_{jkl} + \varepsilon^2 . c_3{}^2 . \underset{(cab)}{C} . \underset{(c)}{E^i}{}_j . \partial_k \underset{(a)}{\varphi} . \partial_l \underset{(b)}{\varphi} \cong \widetilde{R}^i{}_{jkl}$$

*Generators Compatibility Conditions*

These are conditions not on the group itself but on its tensor representation (since they don't affect the structural constants or the resulting algebra). They will be called *Generators Compatibility Conditions*.

For the particular case of a finite single-dimensional transformation (the one associated with a single group generator) the curvature remains invariant up to any order:

(7.3.2) $$\left. \begin{array}{l} U^i{}_j \equiv \delta^i_j + c_3 . \underset{(a)}{E^i}{}_j . \underset{(a)}{\varphi} \\ / \quad \underset{(a)}{\varphi} = (0,..,0,\varphi,0,..,0) \end{array} \right\} \rightarrow \quad \widetilde{R}'^i{}_{jkl} = \widetilde{R}^i{}_{jkl}$$

The *Generators Compatibility Conditions* also allows the curvature to be expressed in a very simple way based on the gauge fields introduced in (7.1.4):

(7.3.3) $$\widetilde{R}^i{}_{jkl} = R^i{}_{jkl} - \Theta^i{}_{jkl} + c_1 . \delta^i_j . F_{kl} - c_3 . \underset{(a)}{E^i}{}_j . \underset{(a)}{H}_{kl}$$

(7.3.4) $$/ \quad \Theta^i{}_{jkl} \equiv -(\nabla_k \Theta^i{}_{jl} - \nabla_l \Theta^i{}_{jk} + \Theta^i{}_{rk} . \Theta^r{}_{jl} - \Theta^i{}_{rl} . \Theta^r{}_{jk})$$ *Matter Curvature*

In terms of the standard connection and structural constants, conditions (7.3.1) can be expressed in the following way:

(7.3.5) $$\widetilde{\nabla}_k \underset{(a)}{E^i{}_j} \equiv 0 \quad \leftrightarrow \quad \nabla_k \underset{(a)}{E^i{}_j} \equiv -c_3.\underset{(abc)}{C}.\underset{(b)}{A_k}.\underset{(c)}{E^i{}_j}+\underset{(a)}{E^i{}_r}.\Theta^r{}_{jk} - \underset{(a)}{E^r{}_j}.\Theta^i{}_{rk}$$

An immediate property emerging from (7.3.1) is the following:

(7.3.6) $$\widetilde{R}^i{}_{rkl}.\underset{(a)}{E^r{}_j} = \widetilde{R}^r{}_{jkl}.\underset{(a)}{E^i{}_r}$$

This can be seen as a commutation rule between (*Gravity – Matter)* curvature-difference and the group generators:

(7.3.7) $$\rightarrow \quad (R^i{}_{rkl} - \Theta^i{}_{rkl}).\underset{(a)}{E^r{}_j} - \underset{(a)}{E^i{}_r}.(R^r{}_{jkl} - \Theta^r{}_{jkl}) = c_3.\underset{(cba)}{C}.\underset{(c)}{E^i{}_j}.\underset{(b)}{H_{kl}}$$

Conditions (7.3.11) are *AT*-invariant with respect to *Electromagnetism* and *Gauge* transformations so they can be introduced into the Lagrangian using invariant constraining fields $\underset{(a)}{\widetilde{C}_{ijk}}$ :

(7.3.8) $$\widetilde{\nabla}_k \underset{(a)}{E^i{}_j} \equiv 0 \quad \rightarrow \quad \widetilde{\nabla}'_k \underset{(a)}{E'^i{}_j} \equiv 0$$

(7.3.9) $$\sqrt{-|g_{\bullet\bullet}|}.g^{kr}.g^{sl}.\underset{(a)}{\widetilde{C}_{kms}}.\widetilde{\nabla}_r \underset{(a)}{E^m{}_l} \qquad / \qquad \underset{(a)}{\widetilde{C}_{ijk}} = -\underset{(a)}{\widetilde{C}_{ikj}} \quad , \quad (\underset{(a)}{\widetilde{C}_{ijk}})'^{(EM)} = (\underset{(a)}{\widetilde{C}_{ijk}})''^{(GF)} = \underset{(a)}{\widetilde{C}_{ijk}}$$

## 7.4 Invariant Torsion Terms

The other Lagrangian terms are those associated with derivatives of the torsion-trace $\partial\hat{\Gamma}_{ij}$. Such term results absolute-invariant for *Electromagnetic* transformations (6.11.1) but not in general for those like (7.1.6). One may be tempted to suppress such terms from the Lagrangian (i.e. by taking $k_2 = k_3 = 0$) but this will exclude *Electromagnetism* from the resulting theory. Because of this the following *EM-Gauge-Compatibility Condition* will be imposed to the gauge-group's transforming parameters in order to grant invariance to the torsion-trace's rotational:

(7.4.1) $$\partial\hat{\Gamma}'_{ij} = \partial\hat{\Gamma}_{ij} \quad \leftrightarrow \quad \underset{(a)}{E^k{}_i}.\partial_k \underset{(a)}{\varphi} \equiv \partial_i\Phi(X)$$ *EM-Gauge-Compatibility Condition*

## 7.5 The Connection's Variational Structure

As mentioned in (2.1.9) the general connection can be decomposed into the sum of independent terms as follows:

(7.5.1) $$\widetilde{\Gamma}^i{}_{jk} = \overline{\Gamma}^i{}_{jk} + \hat{\Gamma}^i{}_{jk}$$

The symmetric connection and the torsion are independent from one another due to their different lower index symmetries and different covariant behavior (connection vs. tensor).

The torsion itself admits an irreducible decomposition in three independent components [9]:

(7.5.2) $$\hat{\Gamma}^i{}_{jk} = \underset{(1)}{\hat{\Gamma}^i{}_{jk}} + \underset{(2)}{\hat{\Gamma}^i{}_{jk}} + \underset{(3)}{\hat{\Gamma}^i{}_{jk}}$$

(7.5.3) $$\underset{(1)}{\hat{\Gamma}^i{}_{jk}} \equiv -\tfrac{1}{3}.(\delta^i_j.\hat{\Gamma}_k - \delta^i_k.\hat{\Gamma}_j)$$

(7.5.4) $$/ \quad \underset{(2)}{\hat{\Gamma}^i{}_{jk}} \equiv \hat{\Gamma}^i{}_{jk} + \hat{\Gamma}_k{}^i{}_j + \hat{\Gamma}_{jk}{}^i$$

(7.5.5) $$\underset{(3)}{\hat{\Gamma}^i{}_{jk}} \equiv \tfrac{1}{3}.(\delta^i_j.\hat{\Gamma}_k - \delta^i_k.\hat{\Gamma}_j) - \hat{\Gamma}_k{}^i{}_j - \hat{\Gamma}_{jk}{}^i$$

By construction the last two have the following properties:

(7.5.6) $\underset{(2)}{\hat{\Gamma}^k}{}_{jk} = 0$

(7.5.7) $\underset{(2)}{\hat{\Gamma}}_{ijk} = -\underset{(2)}{\hat{\Gamma}}_{jik} = -\underset{(2)}{\hat{\Gamma}}_{kji} = -\underset{(2)}{\hat{\Gamma}}_{ikj}$

(7.5.8) $\underset{(3)}{\hat{\Gamma}^k}{}_{jk} = 0$

(7.5.9) $\underset{(3)}{\hat{\Gamma}}_{ijk} + \underset{(3)}{\hat{\Gamma}}_{kij} + \underset{(3)}{\hat{\Gamma}}_{jki} - \underset{(3)}{\hat{\Gamma}}_{jik} - \underset{(3)}{\hat{\Gamma}}_{kji} - \underset{(3)}{\hat{\Gamma}}_{ikj} = 0$

Identities (7.5.7) shows that $\underset{(2)}{\hat{\Gamma}}_{ijk}$ corresponds to the totally skew-symmetric part of the torsion. Both (7.5.6) and (7.5.8) implies that $\underset{(2)}{\hat{\Gamma}^i}{}_{jk}$ and $\underset{(3)}{\hat{\Gamma}^i}{}_{jk}$ are traceless so that all the torsion's trace is contained in $\underset{(1)}{\hat{\Gamma}^i}{}_{jk}$. Since these components are linearly independent because of their symmetries, for the torsion to be null each one has to vanish. That's why the torsion's variations can be decomposed into three independent ones:

(7.5.10) $\delta\hat{\Gamma}^i{}_{jk} = \delta\underset{(1)}{\hat{\Gamma}^i}{}_{jk} + \delta\underset{(2)}{\hat{\Gamma}^i}{}_{jk} + \delta\underset{(3)}{\hat{\Gamma}^i}{}_{jk}$

Finally the connection admits at least four independent variations:

(7.5.11) $\delta\tilde{\Gamma}^i{}_{jk} = \delta\overline{\Gamma}^i{}_{jk} + \delta\underset{(1)}{\hat{\Gamma}^i}{}_{jk} + \delta\underset{(2)}{\hat{\Gamma}^i}{}_{jk} + \delta\underset{(3)}{\hat{\Gamma}^i}{}_{jk}$

Using the metric and the *Group Generators* each of the torsion's components can be written as a function of vector potentials as follows:

(7.5.12) $\underset{(1)}{\hat{\Gamma}^i}{}_{jk} \equiv \tfrac{1}{2}.c_1.(\delta^i_j.A_k - \delta^i_k.A_j)$

(7.5.13) $\underset{(2)}{\hat{\Gamma}^i}{}_{jk} \equiv \tfrac{1}{2}.c_2.\sqrt{-|g_{\bullet\bullet}|}.\varepsilon_{jkrl}.g^{rs}.g^{lm}.(\delta^i_s.B_m - \delta^i_m.B_s)$

(7.5.14) $\underset{(3)}{\hat{\Gamma}^i}{}_{jk} \equiv -\tfrac{1}{6}.c_3.(\delta^i_j.\underset{(a)}{E^r}{}_k.\underset{(a)}{C_r} - \delta^i_k.\underset{(a)}{E^r}{}_j.\underset{(a)}{C_r}) + \tfrac{1}{2}.c_3.(\underset{(a)}{E^i}{}_j.\underset{(a)}{C_k} - \underset{(a)}{E^i}{}_k.\underset{(a)}{C_j} + 2.\underset{(a)}{E}_{jk}.\underset{(a)}{C^i})$

/ $a \in \{1,2,3,4,5,6\}$

The two first vector potentials can be calculated from the torsion by the following expressions:

(7.5.15) $A_i \equiv -\tfrac{2}{3}.c_1{}^{-1}.\delta^r_k.\hat{\Gamma}^k{}_{ir} = -\tfrac{2}{3}.c_1{}^{-1}.\hat{\Gamma}_i$ *Electromagnetic Potential*

(7.5.16) $B_i \equiv -\tfrac{1}{6}.c_2{}^{-1}.\sqrt{-|g^{\bullet\bullet}|}.g_{ik}.g_{rm}.\varepsilon^{krsl}.\hat{\Gamma}^m{}_{sl}$ *Weak Potential*

Taking into account the general form of the $\underset{(3)}{\hat{\Gamma}^i}{}_{jk}$ component it is clear that any torsion can be expressed as a combination of vector potentials associated to the *Lorentz Group* generators:

(7.5.17) $\hat{\Gamma}^i{}_{jk} \equiv -\tfrac{1}{2}.c_3.(\underset{(a)}{E^i}{}_j.\underset{(a)}{C_k} - \underset{(a)}{E^i}{}_k.\underset{(a)}{C_j})$ *Torsion Lorenz Basis*

All the torsion degrees of freedom are then captured by the $\underset{(a)}{C_i}$ vector fields which will be called *Lorentz* potentials. Once they are given, the *Electromagnetic* potential and the *Weak* one can be calculated using expressions (7.5.15) and (7.5.16).

## 7.6 Duality Transformations

On a four-dimensional manifold any totally skew-symmetric part of a tensor of order $r$ can be put into a one-to-one correspondence with its *Dual* of order $(4-r)$ by the following *Duality Transformation*:

(7.6.1) $$e_{ijkl} \equiv \sqrt{-|g_{\bullet\bullet}|}.\varepsilon_{ijkl} \quad , \quad e^{ijkl} \equiv \sqrt{-|g_{\bullet\bullet}|}^{-1}.\varepsilon^{ijkl}$$ *Levi-Civita Tensor*

*Duality Transformations*:

(7.6.2) $$V_i \rightarrow \overset{\circ}{V}_{ijk} \equiv V_r.g^{rs}.e_{sijk}$$

(7.6.3) $$V_{ij} = -V_{ji} \rightarrow \overset{\circ}{V}_{ij} \equiv \tfrac{1}{2}.V_{rs}.g^{ra}.g^{sb}.e_{abij}$$

(7.6.4) $$V_{ijk} = V_{[ijk]} \rightarrow \overset{\circ}{V}_i \equiv \tfrac{1}{6}.V_{krs}.g^{ka}.g^{rb}.g^{sc}.e_{abci}$$

(7.6.5) $$V_{ijkl} = V_{[ijkl]} \rightarrow \overset{\circ}{V} \equiv \tfrac{1}{24}.V_{krsl}.g^{ka}.g^{rb}.g^{sc}.g^{ld}.e_{abcd}$$

The transformation is done by contracting with the Levi-Civita totally skew-symmetric tensor-capacity which was transformed into a tensor by multiplication against the *Reference* density in use. Such transformation can be reversed in the following way:

(7.6.6) $$V_i = \tfrac{1}{6}.g_{ia}.e^{akrs}.\overset{\circ}{V}_{krs}$$

(7.6.7) $$V_{ij} = \tfrac{1}{2}.g_{ia}.g_{jb}.e^{abkr}.\overset{\circ}{V}_{kr} = -\tfrac{1}{2}.\overset{\circ}{V}_{rs}.g^{ra}.g^{sb}.e_{abij} = -\left(\overset{\circ}{V}_{ij}\right)^{\circ}$$

(7.6.8) $$V_{ijk} = \overset{\circ}{V}_a.g_{ir}.g_{jl}.g_{km}.e^{rlma}$$

(7.6.9) $$V_{ijkl} = \overset{\circ}{V}.g_{ia}.g_{jb}.g_{kc}.g_{ld}.e^{abcd}$$

Because such duals can be reversed they contain the same information of their original tensors.
In this paper the following torsion "*dual*" will be considered:

(7.6.10) $$\overset{\circ}{\hat{\Gamma}}{}^i{}_{jk} \equiv \tfrac{1}{2}.\hat{\Gamma}^i{}_{rs}.g^{ra}.g^{sb}.e_{abjk}$$ *Torsion's Dual*

## 7.7 The Torsion's Structural Symmetry

When the *Duality Transformation* is applied to the torsion its structure transforms in such a way that components $\underset{(1)}{\hat{\Gamma}^i{}_{jk}}$ and $\underset{(2)}{\hat{\Gamma}^i{}_{jk}}$ interchange their roles at the time component $\underset{(3)}{\hat{\Gamma}^i{}_{jk}}$ transforms into itself:

(7.7.1) $$\overset{\circ}{\hat{\Gamma}}{}^i{}_{jk} \equiv \left(\hat{\Gamma}^i{}_{jk}\right)^{\circ} \rightarrow \underset{(1)}{\overset{\circ}{\hat{\Gamma}}{}^i{}_{jk}} = \left(\underset{(2)}{\hat{\Gamma}^i{}_{jk}}\right)^{\circ} \quad , \quad \underset{(2)}{\overset{\circ}{\hat{\Gamma}}{}^i{}_{jk}} = \left(\underset{(1)}{\hat{\Gamma}^i{}_{jk}}\right)^{\circ} \quad , \quad \underset{(3)}{\overset{\circ}{\hat{\Gamma}}{}^i{}_{jk}} = \left(\underset{(3)}{\hat{\Gamma}^i{}_{jk}}\right)^{\circ}$$

In terms of the vector potentials this can be expressed as the following interchange-symmetries:

(7.7.2) $$A_i' = -\tfrac{c_2}{c_1}.B_i \quad , \quad B_i' = \tfrac{c_1}{c_2}.A_i$$

So the *Duality Transformation* operates as a symmetry transformation between this pair of vector potentials. What follows is a suggested identification of the previous structure and symmetry with known physical facts.

## 7.8 Forces, Particles and Symmetries of the Standard Model?

Four independent variations in the connection according to section (7.5.11) lead to the concept of four independent forces acting together. When thinking on elementary particles this traduces into four different attributes a particle can have for telling how it interacts with each force-field. These attributes are the *charges* associated to each force. In particle physics every time $N$ particles within a closed system can interchange roles it's because their wave functions conform a multiplet of a given $SU(N)$ symmetry. At the classical level this could correspond to some symmetry taking place between the corresponding vector potentials for which particles act as sources. That's why the following identifications are being made:

- The symmetric-connection variation $\delta\overline{\Gamma}^i{}_{jk}$ will be associated with *mass* as it contributes to the matter stress-energy tensor in equations seen on section (4.5). Through Einstein's equation and the *Compatibility Condition*, *Gravity* is being determined by the *matter* tensor and corresponding contributions of other torsion (gauge) fields.

- Variations on the torsion's trace $\delta\underset{(1)}{\hat{\Gamma}^i{}_{jk}}$ will be associated as shown in section (5.1) with the *Electromagnetic* potential $A_i$ whose rotational generates the Maxwell's fields. *Electric* charges act as sources for those potentials and transmit the *Electromagnetic* force back to those objects exposing them. *Electrons* and *positrons* become particles exposing a pure *Electromagnetic* field which generates only this kind of torsion in it surrounds.

- Variations of the torsion's completely skew-symmetric part $\delta\underset{(2)}{\hat{\Gamma}^i{}_{jk}}$ will be associated to the *Weak* potential $B_i$ with *weak* charges acting as sources and transmitting the associated force back to the containing objects. *Neutrinos* and *antineutrinos* will be particles exposing a pure *Weak* field generating this kind of torsion.

- Variations of the torsion's remaining component $\delta\underset{(3)}{\hat{\Gamma}^i{}_{jk}}$ will be associated to the *Strong* potentials generated by the *color* charges which transmit the *Strong* force back to the containing objects. The three color charges *red*, *green* and *blue* would be associated to the Lorentz generators $\underset{(a)}{X^{+}}{}^i{}_j$ which close under the corresponding $\underset{L^+}{SU(2)}$ algebra. The other three generators $\underset{(a)}{X^{-}}{}^i{}_j$ will be responsible for the associated anti-color charges generating the $\underset{L^-}{SU(2)}$ algebra.

- *Quarks* would be particles having a full torsion contribution where the components $\delta\underset{(1)}{\hat{\Gamma}^i{}_{jk}}$, $\delta\underset{(2)}{\hat{\Gamma}^i{}_{jk}}$ and $\delta\underset{(3)}{\hat{\Gamma}^i{}_{jk}}$ define their attributes (they have the three kind of charges).

- If components $\underset{(1)}{\hat{\Gamma}^i{}_{jk}}$ and $\underset{(2)}{\hat{\Gamma}^i{}_{jk}}$ can be interchanged by *Duality Transformations* then this symmetry would be responsible for the *Electro-Weak* $\underset{D}{SU(2)}$ symmetry between *electrons* and *neutrinos* or *up* and *down quarks*.
- A symmetry (?) related to the cyclic permutation of the Lorentz generators may allow *quarks* to be arranged into $\underset{LG}{SU(3)}$ multiplets of the *Strong Force*.

Although a particle-field modeling is needed for confirming up to the very end this interpretation it looks quite consistent with the *Standard Model* and may give a geometric support at the classical level for what it can be seen at the quantum scale.

Because of this a *quark's* wave function could have a $\underset{EM}{U(1)}\times\underset{D}{SU(2)}\times\underset{LG}{SU(3)}$ gauge symmetry which is considered to be the *Standard Model's* one.

The suggested identifications should be validated lately with a more complete analysis including particle-fields, their spins and most probably a theory based on a complex domain.

## 7.9 The Connection's Decoupled Field Equations

The field equation obtained by a general variation of the connection $\delta\tilde{\Gamma}^i{}_{jk}$ can be decomposed into independent equations when varying against its components. In particular this will be true if the variation is being carried with those vector potentials in play:

(7.9.1) (7.5.11) $$\rightarrow \quad \underset{(\tilde{\Gamma}^\bullet{}_{\bullet\bullet})}{E_i{}^{jk}} = \underset{(\bar{\Gamma}^\bullet{}_{\bullet\bullet})}{E_i{}^{jk}} + \underset{(\hat{\Gamma}^\bullet{}_{\bullet\bullet})}{E_i{}^{jk}} = \underset{(\bar{\Gamma}^\bullet{}_{\bullet\bullet})}{E_i{}^{jk}} + \underset{(\underset{(1)}{\hat{\Gamma}}^\bullet{}_{\bullet\bullet})}{E_i{}^{jk}} + \underset{(\underset{(2)}{\hat{\Gamma}}^\bullet{}_{\bullet\bullet})}{E_i{}^{jk}} + \underset{(\underset{(3)}{\hat{\Gamma}}^\bullet{}_{\bullet\bullet})}{E_i{}^{jk}}$$

(7.9.2) $$\underset{(\tilde{\Gamma}^\bullet{}_{\bullet\bullet})}{E_i{}^{jk}} \equiv 0 \quad \rightarrow \quad \underset{(\bar{\Gamma}^\bullet{}_{\bullet\bullet})}{E_i{}^{jk}} \equiv 0 \;,\; \underset{(\underset{(1)}{\hat{\Gamma}}^\bullet{}_{\bullet\bullet})}{E_i{}^{jk}} \equiv 0 \;,\; \underset{(\underset{(2)}{\hat{\Gamma}}^\bullet{}_{\bullet\bullet})}{E_i{}^{jk}} \equiv 0 \;,\; \underset{(\underset{(3)}{\hat{\Gamma}}^\bullet{}_{\bullet\bullet})}{E_i{}^{jk}} \equiv 0$$

(7.9.3) $$\underset{(\bar{\Gamma}^\bullet{}_{\bullet\bullet})}{E_i{}^{jk}} \equiv 0 \quad \rightarrow \quad \underset{(\tilde{\Gamma}^\bullet{}_{\bullet\bullet})}{E_i{}^{jk}} + \underset{(\tilde{\Gamma}^\bullet{}_{\bullet\bullet})}{E_i{}^{kj}} \equiv 0$$ *Mass Equation*

(7.9.4) $$\underset{(\underset{(1)}{\hat{\Gamma}}^\bullet{}_{\bullet\bullet})}{E_i{}^{jk}} \equiv 0 \quad \rightarrow \quad \underset{(\hat{\Gamma}^\bullet{}_{\bullet\bullet})}{E_r{}^{ri}} \equiv 0$$ *Electromagnetic Equation*

(7.9.5) $$\underset{(\underset{(2)}{\hat{\Gamma}}^\bullet{}_{\bullet\bullet})}{E_i{}^{jk}} \equiv 0 \quad \rightarrow \quad \underset{(\hat{\Gamma}^\bullet{}_{\bullet\bullet})}{\overset{\circ}{E}_r{}^{ri}} \equiv 0$$ *Weak Equation*

(7.9.6) $$\underset{(\underset{(3)}{\hat{\Gamma}}^\bullet{}_{\bullet\bullet})}{E_i{}^{jk}} \equiv 0 \quad \rightarrow \quad \underset{(\hat{\Gamma}^\bullet{}_{\bullet\bullet})}{E_i{}^{jk}} - \underset{(\underset{(1)}{\hat{\Gamma}}^\bullet{}_{\bullet\bullet})}{E_i{}^{jk}} - \underset{(\underset{(2)}{\hat{\Gamma}}^\bullet{}_{\bullet\bullet})}{E_i{}^{jk}} \equiv 0$$ *Strong Equation*

Where a duality transformation was applied to the upper index-pair in (7.9.5) for simplifying the final expression.

With respect to those equations seen in (5.2) and considering the torsion's irreducible components (7.5) this decomposition establish the following equation-equivalences under the *Relativistic Conformal Gauge* (5.2.10) and the *Generalized Compatibility Condition* (5.2.11):

(7.9.7) $$\underset{(\hat{\Gamma}^{\bullet}{}_{\bullet\bullet})_{(1)}}{E_i{}^{jk}} \equiv 0 \qquad \rightarrow \qquad \bar{Q}^i = 0 \qquad / \qquad \tilde{Q}^i = 0$$

(7.9.8) $$\underset{(\hat{\Gamma}^{\bullet}{}_{\bullet\bullet})_{(2)}}{E_i{}^{jk}} \equiv 0 \qquad \rightarrow \qquad 2.\overline{\overline{\tilde{R}}}.\underset{(2)}{\hat{\Gamma}^i{}_{jk}} = (Q_{jk}{}^i + Q^i{}_{jk} + Q_k{}^i{}_j)$$

(7.9.9) $$\underset{(\hat{\Gamma}^{\bullet}{}_{\bullet\bullet})_{(3)}}{E_i{}^{jk}} \equiv 0 \qquad \rightarrow \qquad 2.\overline{\overline{\tilde{R}}}.\underset{(3)}{\hat{\Gamma}^i{}_{jk}} = \tfrac{1}{3}.(\delta^i_j.\bar{Q}_k - \delta^i_k.\bar{Q}_j) - Q_k{}^i{}_j - Q^i{}_{jk}$$

## 7.10 Holotrinos: the Fundamental Particles?

The *Electromagnetic* field was identified with the first torsion component $\underset{(1)}{\hat{\Gamma}^i{}_{jk}}$ which would be the whole torsion generated by an object exposing only and *electric* charge:

(7.10.1) $$\hat{\Gamma}^i{}_{jk} = \tfrac{1}{2}.c_1.(\delta^i_j.A_k - \delta^i_k.A_j)$$

According to (7.5.17) such torsion can be expressed as a combination of *Lorentz* generators and potentials. For defining a general *Electromagnetic* potential $A_i$ the *Lorentz* ones should have the following form:

(7.10.2) $$\tfrac{1}{2}.c_1.(\delta^i_j.A_k - \delta^i_k.A_j) = \tfrac{1}{2}.c_3.(\underset{(a)}{E^i{}_j}.\underset{(a)}{C_k} - \underset{(a)}{E^i{}_k}.\underset{(a)}{C_j})$$

$$\leftrightarrow \begin{cases} \underset{(1)}{C_i} = \frac{c_1}{c_3}.(A_1, A_0, 0, 0) \;, & \underset{(2)}{C_i} = \frac{c_1}{c_3}.(A_2, 0, A_0, 0) \;, & \underset{(3)}{C_i} = \frac{c_1}{c_3}.(A_3, 0, 0, A_0) \\ \underset{(4)}{C_i} = \frac{c_1}{c_3}.(0, 0, -A_3, A_2) \;, & \underset{(5)}{C_i} = \frac{c_1}{c_3}.(0, A_3, 0, -A_1) \;, & \underset{(6)}{C_i} = \frac{c_1}{c_3}.(0, -A_2, A_1, 0) \end{cases}$$

Following the generator definition given in (7.2.5) these *Lorentz* potentials can be more suitably renamed as:

(7.10.3) $$\begin{cases} \underset{(1)}{C_i} \rightarrow \underset{(x)}{k_i} \;, & \underset{(2)}{C_i} \rightarrow \underset{(y)}{k_i} \;, & \underset{(3)}{C_i} \rightarrow \underset{(z)}{k_i} \\ \underset{(4)}{C_i} \rightarrow \underset{(x)}{l_i} \;, & \underset{(5)}{C_i} \rightarrow \underset{(y)}{l_i} \;, & \underset{(6)}{C_i} \rightarrow \underset{(z)}{l_i} \end{cases}$$

Then the equality (7.10.2) would read:

(7.10.4) $$\tfrac{1}{2}.c_1.(\delta^i_j.A_k - \delta^i_k.A_j) = \tfrac{1}{2}.c_3.(\underset{(a)}{K^i{}_j}.\underset{(a)}{k_k} - \underset{(a)}{K^i{}_k}.\underset{(a)}{k_j}) + \tfrac{1}{2}.c_3.(\underset{(a)}{L^i{}_j}.\underset{(a)}{l_k} - \underset{(a)}{L^i{}_k}.\underset{(a)}{l_j})$$

$$/ \qquad a \in \{x, y, z\}$$

If each of the *Lorentz* vector potentials can be associated to an elementary particle then those particles should be more fundamental than *leptons* and *quarks*.

They would be called *Holotrinos* (from ***Ho****-mogenous* ***Lo****-rentz* group) and could be arranged in two sets of three similar particles called *Acceletrinos* and *Rotatrinos* (from *Acceleration* and *Rotation*):

(7.10.5) $k^x, k^y, k^z \rightarrow \quad Acceletrinos \quad , \qquad l^x, l^y, l^z \rightarrow \quad Rotatrinos$

For an electrically charged object at rest the *Electromagnetic* potential simplifies to (*electrostatic* case):

(7.10.6) $A_i = (A_0(X),0,0,0)$

If the object is a single *electron* then in the *Lorentz* base the *holotronic* potentials should be:

(7.10.7) $\underset{(x)}{k_i} = \frac{c_1}{c_3}.(0, A_0,0,0) \quad , \quad \underset{(y)}{k_i} = \frac{c_1}{c_3}.(0,0, A_0,0) \quad , \quad \underset{(z)}{k_i} = \frac{c_1}{c_3}.(0,0,0, A_0)$

An *electron* will then be composed of three different *acceletrinos*. Since they contribute equally to determine the final potential they should have the same electric charge which then becomes $-\frac{1}{3}.e$. *Holotrinos* would then combine to give *electrons*, *neutrinos*, *down* and *up quarks*. This means *leptons* and *quarks* could be taken to the same footing becoming clearer their interaction as a *particle family*.

Taking into account *Dual* symmetries, the final *electric* and *weak* charges and the fact that the combined spin should be $\frac{1}{2}.\hbar$ (considering *holotrinos* as fermions of spin $\frac{1}{2}.\hbar$) the know particles would admit the following *electro-weak-static* composition in terms of these fundamental particles:

(7.10.8) $e^- = \underset{(x)}{k}(0,-1,0,0) + \underset{(y)}{k}(0,0,-1,0) + \underset{(z)}{k}(0,0,0,-1)$ *Electron*

(7.10.9) $\nu_e = \underset{(x)}{l}(0,-1,0,0) + \underset{(y)}{l}(0,0,-1,0) + \underset{(z)}{l}(0,0,0,-1)$ *Electron-Neutrino*

(7.10.10) $d_R = \underset{(x)}{k}(0,-1,0,0) + \underset{(y)}{l}(0,0,-1,0) + \underset{(z)}{l}(0,0,0,-1)$ *Red Down-Quark*

(7.10.11) $d_G = \underset{(x)}{l}(0,-1,0,0) + \underset{(y)}{k}(0,0,-1,0) + \underset{(z)}{l}(0,0,0,-1)$ *Green Down-Quark*

(7.10.12) $d_B = \underset{(x)}{l}(0,-1,0,0) + \underset{(y)}{l}(0,0,-1,0) + \underset{(z)}{k}(0,0,0,-1)$ *Blue Down-Quark*

(7.10.13) $u_R = \underset{(x)}{l}(0,+1,0,0) + \underset{(y)}{k}(0,0,+1,0) + \underset{(z)}{k}(0,0,0,+1)$ *Red Up-Quark*

(7.10.14) $u_G = \underset{(x)}{k}(0,+1,0,0) + \underset{(y)}{l}(0,0,+1,0) + \underset{(z)}{k}(0,0,0,+1)$ *Green Up-Quark*

(7.10.15) $u_B = \underset{(x)}{k}(0,+1,0,0) + \underset{(y)}{k}(0,0,+1,0) + \underset{(z)}{l}(0,0,0,+1)$ *Blue Up-Quark*

Where the coupling constants and potential units were adjusted for leaving the *electron* value $-1$ as the unit reference.

When *quarks* combine for integrating *protons* and *neutrons* their *electro-weak-static* composition results opposite to those for *electrons* and *neutrinos* respectively:

(7.11.16) $p^+ = \underset{(x)}{k}(0,+1,0,0) + \underset{(y)}{k}(0,0,+1,0) + \underset{(z)}{k}(0,0,0,+1)$ *Proton*

(7.11.17) $n = \underset{(x)}{l}(0,+1,0,0) + \underset{(y)}{l}(0,0,+1,0) + \underset{(z)}{l}(0,0,0,+1)$ *Neutron*

While *acceletrinos* would be responsible for the final *electric* charge the *rotatrinos* would do the same for the *weak* charge.

## 8 Conformal Gauge Field Gravitation

When the torsion is allowed to have any value more degrees of freedom are available to introduce what will correspond to new gauge field potentials into the theory. The corresponding gauge invariance results from the absolute invariance associated to the loop operators in (7.1.6) which depend on the *Group Generators* fields and the corresponding phase-function parameters. The *Group Generators* are new geometrical variables that have to meet the *Metric Constraint* (7.2.1) and the *Generators Compatibility Conditions* (7.3.1) along with the usual gauge algebra (7.1.1) in order to fit into the theory and preserve standard measurements (i.e. intervals) and loop invariance. They are six covariant skew-symmetric tensor fields adding the Lie's algebra of the *Homogenous Lorentz Group* $SU(2)\times SU(2)$ to the already existing for the *Conformal Electromagnetic Gravity* represented here as $GR\times U(1)$. The phase-function parameters are constrained by the *EM-Gauge-Compatibility Condition* (7.4.1) for preserving *Electromagnetism*.

### 8.1 Conformal Gauge Field Gravity Action

The *Conformal Gauge Field Gravity* action will be the *Conformal Electromagnetic Gravity* one with those extra terms necessary for introducing the *Group Generators* and their properties.
Although torsion is relaxed from compatibility condition (4.5.5) for allowing new gauge fields, compatibility with *General Relativity* still requires the delta-condition (4.5.6) to hold. This is equivalent to the *Generalized Compatibility Condition* (5.2.11) which is *AT*-invariant both to *Electromagnetic* and *Gauge* transformations so it can be granted as a field equation by introducing the following Lagrangian term:

(8.1.1)
$$\sqrt{-|g_{\bullet\bullet}|}.g^{kr}.\tilde{C}_{kls}.(\tilde{\nabla}_r g^{ls} - g^{ls}.g_{ro}.\tilde{\nabla}_p g^{op}) \qquad / \qquad (\tilde{C}_{ijk})'^{(EM)} = (\tilde{C}_{ijk})''^{(GF)} = \tilde{C}_{ijk}$$

This condition may have been introduced in the previous *Conformal Electromagnetic Gravity* Lagrangian but that was avoided only for showing that a compatible theory was possible having a standard solution set bigger than the known for *General Relativity + Electromagnetism* (it is bigger since it includes standard non-compatible solutions as well).

After adding the term (8.1.1) and those for ensuring the generators properties (7.1.1), (7.2.2) and (7.3.1), the gauge unification action results:

(8.1.2) *Conformal Gauge Field Gravity Action*

$$\begin{aligned} I_{CGFG} \equiv k_0.\int_D \sqrt{-|g_{\bullet\bullet}|}.g^{kr}.g^{sl}.\Big( & \overline{\overline{\tilde{R}}}_{kr}.\overline{\overline{\tilde{R}}}_{sl} + k_1.(\hat{\overline{\tilde{R}}}_{ks} - \tfrac{1}{2}.\overline{\tilde{R}}_{ks}).(\hat{\overline{\tilde{R}}}_{rl} - \tfrac{1}{2}.\overline{\tilde{R}}_{rl}) + k_2.\partial\hat{\Gamma}_{ks}.\partial\hat{\Gamma}_{rl} + k_3.(\hat{\overline{\tilde{R}}}_{ks} - \tfrac{1}{2}.\overline{\tilde{R}}_{ks}).\partial\hat{\Gamma}_{rl} + \\ & + k_4.g^{mn}.g_{op}.(\tilde{R}^o{}_{ksm} + \tilde{R}^o{}_{mks} + \tilde{R}^o{}_{smk}).(\tilde{R}^p{}_{rln} + \tilde{R}^p{}_{nrl} + \tilde{R}^p{}_{lnr}) + \\ & - (\tfrac{1}{2}.k_1 + \tfrac{3}{8}.k_3 + 3.k_4).(\tilde{R}^m{}_{kso} + \tilde{R}^m{}_{oks} + \tilde{R}^m{}_{sok}).(\tilde{R}^o{}_{rlm} + \tilde{R}^o{}_{mrl} + \tilde{R}^o{}_{lmr}) + \\ & + \tfrac{1}{4}.k_5.g_{sl}.\tilde{C}_{kmn}.(\tilde{\nabla}_r g^{mn} - g^{mn}.g_{ro}.\tilde{\nabla}_p g^{op}) + k_6.\underset{(a)}{\tilde{C}_{krsm}}.\underset{(a)}{E^m{}_l} + k_7.\underset{(a)}{\tilde{C}_{kms}}.\tilde{\nabla}_r \underset{(a)}{E^m{}_l} + \\ & + k_8.\underset{(ab)}{\tilde{C}_{krsm}}.(\underset{(a)}{E^m{}_o}.\underset{(b)}{E^o{}_l} - \underset{(b)}{E^m{}_o}.\underset{(a)}{E^o{}_l} - \underset{(cab)}{C}.\underset{(c)}{E^m{}_l})\Big).d\Omega \end{aligned}$$

(8.1.3)
$$/ \quad \begin{cases} 3.k_2 + k_3 \equiv \frac{32.k_g.\lambda_c}{3.c^4.c_1{}^2} \quad , \quad \tilde{C}_{ijk} \equiv \tilde{C}_{ikj} \\ \underset{(a)}{\tilde{C}_{ijkl}} \equiv \underset{(a)}{\tilde{C}_{jikl}} \equiv \underset{(a)}{\tilde{C}_{ijlk}} \quad , \quad \underset{(a)}{\tilde{C}_{ijk}} \equiv -\underset{(a)}{\tilde{C}_{ikj}} \quad , \quad \underset{(ab)}{\tilde{C}_{ijkl}} \equiv \underset{(ab)}{\tilde{C}_{jikl}} \equiv -\underset{(ab)}{\tilde{C}_{ijlk}} \equiv -\underset{(ba)}{\tilde{C}_{ijkl}} \\ \underset{(cab)}{C} \equiv -\underset{(cba)}{C} \quad , \quad \underset{(eda)}{C}.\underset{(dbc)}{C} + \underset{(edc)}{C}.\underset{(dab)}{C} + \underset{(edb)}{C}.\underset{(dca)}{C} \equiv 0 \end{cases}$$

## 8.2 Field Equations

The corresponding field equations become:

(8.2.1) $\delta\tilde{C}_{kmn}\ )\qquad \tilde{\nabla}_i g^{jk} - g^{jk}.g_{ir}.\tilde{\nabla}_s g^{rs} = 0$

(8.2.2) $\delta\underset{(a)}{\tilde{C}_{krsm}}\ )\qquad \underset{(a)}{E_{ij}} = -\underset{(a)}{E_{ji}}$

(8.2.3) $\delta\underset{(a)}{\tilde{C}_{kms}}\ )\qquad \tilde{\nabla}_k \underset{(a)}{E^i{}_j} = 0$

(8.2.4) $\delta\underset{(ab)}{\tilde{C}_{krsm}}\ )\qquad \underset{(a)}{E^i{}_k}.\underset{(b)}{E^k{}_j} - \underset{(b)}{E^i{}_k}.\underset{(a)}{E^k{}_j} = \underset{(cab)}{C}.\underset{(c)}{E^i{}_j}$

(8.2.5) $\delta\underset{(a)}{E^m{}_n}\ )\qquad -k_6.\underset{(a)}{\tilde{C}^r{}_r{}^i{}_j} + k_7.\tilde{\Pi}_k\underset{(a)}{\tilde{C}^k{}_j{}^i} - 2.k_8.\left(\underset{(b)}{E^i{}_k}.\underset{(ab)}{\tilde{C}^r{}_r{}^k{}_j} - \underset{(b)}{E^k{}_j}.\underset{(ab)}{\tilde{C}^r{}_r{}^i{}_k} - \tfrac{1}{2}.\underset{(cb)}{\tilde{C}^r{}_r{}^i{}_j}.\underset{(acb)}{C}\right) = 0$

(8.2.6) $\delta g^{ij}\ )\ \overline{\overleftrightarrow{\tilde{R}}}.(\overline{\overleftrightarrow{\tilde{R}}}_{ij} - \tfrac{1}{4}.\overline{\overleftrightarrow{\tilde{R}}}.g_{ij}) =$

$$= k_1.(-(\hat{\overleftrightarrow{\tilde{R}}}_i{}^k - \tfrac{1}{2}.\overleftarrow{\tilde{R}}_i{}^k).(\hat{\overleftrightarrow{\tilde{R}}}_{jk} - \tfrac{1}{2}.\overleftarrow{\tilde{R}}_{jk}) + \tfrac{1}{4}.(\hat{\overleftrightarrow{\tilde{R}}}^{kr} - \tfrac{1}{2}.\overleftarrow{\tilde{R}}^{kr}).(\hat{\overleftrightarrow{\tilde{R}}}_{kr} - \tfrac{1}{2}.\overleftarrow{\tilde{R}}_{kr}).g_{ij}) +$$

$$+ k_2.(-\partial\hat{\Gamma}_i{}^k.\partial\hat{\Gamma}_{jk} + \tfrac{1}{4}.\partial\hat{\Gamma}^{kr}.\partial\hat{\Gamma}_{kr}.g_{ij}) +$$

$$+ \tfrac{1}{2}.k_3.(-(\hat{\overleftrightarrow{\tilde{R}}}_i{}^k - \tfrac{1}{2}.\overleftarrow{\tilde{R}}_i{}^k).\partial\hat{\Gamma}_{jk} - (\hat{\overleftrightarrow{\tilde{R}}}_j{}^k - \tfrac{1}{2}.\overleftarrow{\tilde{R}}_j{}^k).\partial\hat{\Gamma}_{ik} + \tfrac{1}{2}.(\hat{\overleftrightarrow{\tilde{R}}}^{kr} - \tfrac{1}{2}.\overleftarrow{\tilde{R}}^{kr}).\partial\hat{\Gamma}_{kr}.g_{ij}) +$$

$$+ \tfrac{1}{2}.k_4.\Big(-3.(\tilde{R}_{pi}{}^{\ln} + \tilde{R}_p{}^n{}_i{}^l + \tilde{R}_p{}^{\ln}{}_i).(\tilde{R}^p{}_{j\ln} + \tilde{R}^p{}_{njl} + \tilde{R}^p{}_{\ln j}) +$$

$$+ (\tilde{R}_i{}^{r\ln} + \tilde{R}_i{}^{nrl} + \tilde{R}_i{}^{\ln r}).(\tilde{R}_{jr\ln} + \tilde{R}_{jnrl} + \tilde{R}_{j\ln r}) +$$

$$+ \tfrac{1}{2}.(\tilde{R}_p{}^{r\ln} + \tilde{R}_p{}^{nrl} + \tilde{R}_p{}^{\ln r}).(\tilde{R}^p{}_{r\ln} + \tilde{R}^p{}_{nrl} + \tilde{R}^p{}_{\ln r}).g_{ij}\Big) +$$

$$-(\tfrac{1}{2}.k_1 + \tfrac{3}{8}.k_3 + 3.k_4).\Big(-(\tilde{R}^m{}_i{}^l{}_o + \tilde{R}^m{}_{oi}{}^l + \tilde{R}^{ml}{}_{oi}).(\tilde{R}^o{}_{jlm} + \tilde{R}^o{}_{mjl} + \tilde{R}^o{}_{lmj}) +$$

$$+ \tfrac{1}{4}.(\tilde{R}^{mrl}{}_o + \tilde{R}^m{}_o{}^{rl} + \tilde{R}^{ml}{}_o{}^r).(\tilde{R}^o{}_{rlm} + \tilde{R}^o{}_{mrl} + \tilde{R}^o{}_{lmr}).g_{ij}\Big)$$

(8.2.7) $\delta\tilde{\Gamma}^i{}_{jk}\Big)\ \tilde{\Pi}_r\partial\hat{\Gamma}^{ri} + \hat{\Gamma}^i{}_{sr}.\partial\hat{\Gamma}^{sr} = 0$

(8.2.8) $\delta\tilde{\Gamma}^i{}_{jk}\Big)\ \tilde{\Pi}_i(\overline{\overleftrightarrow{\tilde{R}}}.g^{jk}) - \delta_i^k.\tilde{\Pi}_r(\overline{\overleftrightarrow{\tilde{R}}}.g^{jr}) + 2.\hat{\Gamma}^k{}_{ir}.(\overline{\overleftrightarrow{\tilde{R}}}.g^{jr}) =$

$$= -k_1.(\tilde{\Pi}_i(\hat{\overleftrightarrow{\tilde{R}}}^{jk} - \tfrac{1}{2}.\overleftarrow{\tilde{R}}^{jk}) - \delta_i^k.\tilde{\Pi}_r(\hat{\overleftrightarrow{\tilde{R}}}^{jr} - \tfrac{1}{2}.\overleftarrow{\tilde{R}}^{jr}) + 2.\hat{\Gamma}^k{}_{ir}.(\hat{\overleftrightarrow{\tilde{R}}}^{jr} - \tfrac{1}{2}.\overleftarrow{\tilde{R}}^{jr})) +$$

$$+ k_1.\delta_i^j.(\tilde{\Pi}_p(\hat{\overleftrightarrow{\tilde{R}}}^{pk} - \tfrac{1}{2}.\overleftarrow{\tilde{R}}^{pk}) + \hat{\Gamma}^k{}_{rp}.(\hat{\overleftrightarrow{\tilde{R}}}^{rp} - \tfrac{1}{2}.\overleftarrow{\tilde{R}}^{rp})) +$$

$$-\tfrac{1}{2}.k_3.(\tilde{\Pi}_i\partial\hat{\Gamma}^{jk} - \delta_i^k.\tilde{\Pi}_p\partial\hat{\Gamma}^{jp} + 2.\hat{\Gamma}^k{}_{ip}.\partial\hat{\Gamma}^{jp}) +$$

$$-\tfrac{1}{2}.k_3.\delta_i^k.(\tilde{\Pi}_p(\hat{\overleftrightarrow{\tilde{R}}}^{pj} - \tfrac{1}{2}.\overleftarrow{\tilde{R}}^{pj}) + \hat{\Gamma}^j{}_{pr}.(\hat{\overleftrightarrow{\tilde{R}}}^{pr} - \tfrac{1}{2}.\overleftarrow{\tilde{R}}^{pr})) +$$

$$+\tfrac{1}{2}.k_3.\delta_i^j.(\tilde{\Pi}_p(\hat{\overleftrightarrow{\tilde{R}}}^{pk} - \tfrac{1}{2}.\overleftarrow{\tilde{R}}^{pk}) + \hat{\Gamma}^k{}_{pr}.(\hat{\overleftrightarrow{\tilde{R}}}^{pr} - \tfrac{1}{2}.\overleftarrow{\tilde{R}}^{pr})) +$$

$$+ 2.(\tfrac{1}{2}.k_1 + \tfrac{3}{8}.k_3 + 3.k_4).(\tilde{\Pi}_p(\tilde{R}^{kjp}{}_i + \tilde{R}^k{}_i{}^{jp} + \tilde{R}^{kp}{}_i{}^j) + \hat{\Gamma}^k{}_{rp}.(\tilde{R}^{pjr}{}_i + \tilde{R}^p{}_i{}^{jr} + \tilde{R}^{pr}{}_i{}^j) +$$

$$+ \tilde{\Pi}_p(\tilde{R}^{jpk}{}_i + \tilde{R}^j{}_i{}^{pk} + \tilde{R}^{jk}{}_i{}^p) + \hat{\Gamma}^k{}_{rp}.(\tilde{R}^{jrp}{}_i + \tilde{R}^j{}_i{}^{rp} + \tilde{R}^{jp}{}_i{}^r) +$$

$$+ \tilde{\Pi}_p(\tilde{R}^{pkj}{}_i + \tilde{R}^p{}_i{}^{kj} + \tilde{R}^{pj}{}_i{}^k) + \hat{\Gamma}^k{}_{rp}.(\tilde{R}^{rpj}{}_i + \tilde{R}^r{}_i{}^{pj} + \tilde{R}^{rj}{}_i{}^p)) +$$

$$-6.k_4.(\tilde{\Pi}_p(\tilde{R}_i{}^{jpk} + \tilde{R}_i{}^{kjp} + \tilde{R}_i{}^{pkj}) + \hat{\Gamma}^k{}_{rp}.(\tilde{R}_i{}^{jrp} + \tilde{R}_i{}^{pjr} + \tilde{R}_i{}^{rpj})) +$$

$$-\tfrac{1}{2}.k_5.(-\tilde{C}^k{}_i{}^j - \tilde{C}^{kj}{}_i + \tilde{C}_i{}^r{}_r.g^{jk} + \tilde{C}^{jr}{}_r.\delta_i^k) +$$

$$-\tfrac{1}{2}.k_7.(-\underset{(a)}{\tilde{C}^k{}_i{}^r}.\underset{(a)}{E^j{}_r} + \underset{(a)}{\tilde{C}^k{}_r{}^j}.\underset{(a)}{E^r{}_i}) = 0$$

The connection-equation coming from variation $\delta\widetilde{\Gamma}^i{}_{jk}$ was opened in two in order to simplify the resulting expressions. Others can be similarly extracted out of (8.2.8) according to (7.9).

Notice that the group generators and the constraining variables don't add new terms to the stress-energy equation (8.2.6). All their influence on this equation is being done through the way they integrate and define the connection.

## 8.3 GR-Compatible Field Equations

With the following *Gauge-Field Gravity Connection* solutions can be found for the previous equation set, which under the *Relativistic Conformal Gauge* becomes:

(8.3.1) $$\overline{\overline{\widetilde{R}}} = 4.\lambda_c$$

(8.3.2) $$\widetilde{\Gamma}^i{}_{jk} \equiv \Gamma^i{}_{jk} + \Theta^i{}_{jk} + c_1.\delta^i_j.A_k - c_3.\underset{(a)}{E^i{}_j}.\underset{(a)}{A_k}$$ *GF-GR Connection*

(8.3.3) $$/ \quad \Theta^i{}_{jk} \equiv \Theta^i{}_{kj} \quad , \quad \Theta_i{}^k{}_j + \Theta^k{}_{ij} - \delta^k_i.(\Theta_j{}^r{}_r + \Theta^r{}_{jr}) \equiv 0$$

(8.3.4) $$\nabla_i g^{jk} = 0$$

(8.3.5) $$A'_i \equiv A_i + \tfrac{1}{3}.\tfrac{c_3}{c_1}.\underset{(a)}{E^k{}_i}.\underset{(a)}{A_k}$$

(8.3.6) $$F'_{ij} \equiv \partial_i A'_j - \partial_j A'_i$$

(8.3.7) $$\nabla_k F'^{ik} = 0$$

(8.3.8) $$\partial_k F'_{ij} + \partial_j F'_{ki} + \partial_i F'_{jk} = 0$$

(8.3.9) $$R_{ij} - \tfrac{1}{2}.R.g_{ij} = \tfrac{8.\pi.k_g}{c^4}.T_{ij} - \lambda_c.g_{ij} + \tfrac{8.\pi.k_g}{c^4}.\Big(\tfrac{1}{4.\pi}.(-F'^k_i.F'_{jk} + \tfrac{1}{4}.F'^{kr}.F'_{kr}.g_{ij})\Big) + (other_terms)...$$

(8.3.10) $$T_{ij} \equiv \tfrac{c^4}{8.\pi.k_g}.\Big(\overline{\overline{\Theta}}_{ij} - \tfrac{1}{2}.g^{kr}.\overline{\overline{\Theta}}_{kr}.g_{ij}\Big)$$

/

(8.3.11) $$\overline{\overline{\Theta}}_{ij} \equiv -\nabla_k\Theta^k{}_{ij} + \Theta^k{}_{ir}.\Theta^r{}_{jk} + \tfrac{1}{2}.(\nabla_i\Theta^k{}_{jk} + \nabla_j\Theta^k{}_{ik}) - \Theta^r{}_{ij}.\Theta^k{}_{rk}$$

(8.3.12) + *(other equations)...*

Terms in (8.3.6) and equations in (8.3.11) were contained for remarking the resulting ones.
This is again *General Relativity* having contributions from a matter field $T_{ij}$, *Electromagnetism*, other *Gauge Fields* and those auxiliary geometric fields needed for the theory's consistency $\widetilde{C}_{ijk}$, $\underset{(a)}{\widetilde{C}_{ijkl}}$, $\underset{(a)}{\widetilde{C}_{ijk}}$ and $\underset{(ab)}{\widetilde{C}_{ijkl}}$.

The *Electromagnetic* potential $A'_i$ is a combination of a generic one $A_i$ and the corresponding contribution coming from the *Lorentz* potentials.

## 9 Conclusions

The present theory achieves unification between *Gravitation*, *Electromagnetism* and other *Gauge Fields* at a classical level on a real four dimensional manifold. Such dimensionality allows a metric's conformal symmetry which combined with the connection's lambda invariance enable introducing the $U(1)$ *Electromagnetic* gauge symmetry through a new transformation called *Absolute Loop Transformation*.

A *Geometrical Postulate* was given for building the Lagrangian based on fundamental geometric objects which characterize the manifold: the metric, the generalized connection and the *Gauge Group's Generators*. Other fields were introduced solely for imposing symmetries and differential constraints on the previous ones. By this all the terms in the Lagrangian become carriers for some geometric specification.

One of the main symmetries in the theory was identified as the *Absolute Loop Transformation* which leaves the connection's equivalent class invariant under its action. While being quite particular it allows introducing the gauge group $U(1)\times SU(2)\times SU(2)$ which may be a projection of the *Standard Model's* one $U(1)\times SU(2)\times SU(3)$ seen at the particle level since the following chain exists: $SU(3)\supset SU(2)$.

For the $U(1)$ gauge corresponding to *Electromagnetism* it was mandatory for the theory to support the metric's *Conformal Symmetry* in order to free this fundamental tensor and the involved geometry from the influence of an *Electromagnetic* regauging. This symmetry demanded the manifold to be four-dimensional. Because the remaining $SU(2)\times SU(2)$ comes from the *Homogeneous Lorentz Group* any regauging naturally leaves the metric invariant.

Different gauge potentials were introduced through independent connection components having no relation with the metric. In this sense this theory doesn't follow the Kaluza-Klein construct or need any dimensional compactification to reach the space-time dimensionality perceived.

The theory was developed in order to contain *General Relativity's* solutions. Compatibility is achieved by choosing a given conformal gauge and ensuring the connection meets some internal properties. When this happens the field equations are found to be those of *General Relativity* with a stress-energy tensor that can model continuous-matter fields and contains the additional gauge fields introduced.

Continuous-matter can be introduced through the *symmetric-Delta* tensor in the connection so a geometric origin can be finally associated to it. If all kind of matter can be modeled following this construction and the corresponding thermo-dynamic considerations that is something to be found in future analysis.

Under compatibility a non-null cosmological constant is needed. For solutions belonging to the *Compact Set* such constant becomes null. Continuity between those two solution-types can be achieved if the *Relativistic Conformal Gauge* is abandoned in the transition zone (non-compatible solutions). This shows that although different physic laws are applied when describing compatible and compact solutions some linkage may exist between them. This opens the door for finding out what happens to the physical information when relativistic solutions reach extreme limits (like in the core of a black hole or at the initial moments of the Big-Bang) because non-compatible and compact solutions can be used for describing the continuous evolution of those compatible ones when *General Relativity* cease to be valid. This would be the case if a classical theory is still adequate for describing reality at those events.

The group generators were introduced as new geometrical fields. This was done in such a way that all their usual properties along with those differential conditions needed were determined by the theory itself.

Gauge fields were introduced through the torsion tensor. Independent internal components in the torsion were expressed using the corresponding vector potentials and gauge group generators. Based on them the torsion may expose symmetries like the duality transformation which along with its internal structure can be used for understanding those particles contained in a family and forces acting between them. This suggests that in the case the theory can be extended to represent particle fields there is a chance the torsion symmetries turn out to be a geometric justification for the existing model of particles and interactions of the *Standard Model*.

The actual formulation does not allow introducing currents in the standard way. This is regarded as a demand for the missing particle fields mentioned previously. It is understood that particle fields should be introduced without violating the *Geometrical Postulate* so currents have to be composed of fundamental variables (existing or new ones).

On section (4.4) it was seen that uniqueness on the solutions of the *Curvature-Phase* equation demanded some constraint to be met for calculating the *Global Phase*. This lead to the concept of an associated family of curves (*Path Dependent*) or points (*Path Independent*) which become related to them. On section (4.6) it was pointed out that this also was the case when the *Relativistic Conformal Gauge* was considered. The relation between solutions, family of curves and compatibility strongly suggest that it may be possible the existence of elementary well defined solutions based on oscillating curve-sections which evolve consistently under *General Relativity's* conditions. These are no others but *Strings.* Based on collections of such elementary solutions bigger ones could be obtained. In that case the image of *Strings* being the foundation for all know particles would be justified. Since the actual manifold is constrained to four dimensions it is not possible to claim that *Strings* can be completely defined in it since the renormalization condition on a quantum-field theory demands the containing manifold to have 10 or 26 dimensions. This becomes an additional motivation for extending the actual theory to be part of a bigger one where particles can be introduced and the *Internal Space* justified.